\documentclass[preprint,12pt]{elsarticle}

\usepackage{amssymb}

\usepackage{graphicx}
\usepackage{subfigure}
\usepackage{dcolumn}
\usepackage{bm}
\usepackage{enumitem}
\usepackage{amsmath,mathrsfs,amsfonts,dsfont,bm}
\usepackage{hyperref}
\hypersetup{colorlinks=true, citecolor=blue, urlcolor=blue, linkcolor=blue}

\journal{Applied Mathematics and Computation}

\begin{document}

\begin{frontmatter}



\title{Emergence of cooperation in a population with bimodal response behaviors}




\author[1,3]{Lin Ma}
\author[2]{Jiqiang Zhang}
\author[1]{Guozhong Zheng}
\author[1,3]{Rizhou Liang}
\author[1]{Li Chen\corref{cor1}}
\ead{chenl@snnu.edu.cn}
\cortext[cor1]{Corresponding Author}
\address[1]{School of Physics and Information Technology, Shaanxi Normal University, Xi'an 710061, P. R. China}
\address[2]{School of Physics and Electronic-Electrical Engineering, Ningxia University, Yinchuan 750021, P. R. China}
\address[3]{School of Systems Science, Beijing Normal University, Beijing 100875,  P. R. China}

\begin{abstract}
We human beings show remarkable adaptability in response to complex surroundings, we adopt different behavioral modes at different occasions, such response multimodality is critical to our survival. Yet, how this behavioral multimodality affects the evolution of cooperation remains largely unknown. 
Here we build a toy model to address this issue by considering a population with bimodal response behaviors, or specifically, with the Fermi and Tit-for-tat updating rules. While the former rule tends to imitate the strategies of those neighbors who are doing well, the latter repeats what their neighbors did to them. In a structural mixing implementation, where the updating rule is fixed for each individual, we find that a moderate mode mixture unexpectedly boosts the overall cooperation level of the population. The boost is even more pronounced in the probabilistic mixing, where each individual randomly chooses one of the two modes at each step, and full cooperation is seen in a wide range. 
These findings are robust to the underlying topology of the population. Our mean-field treatment reveals that the cooperation prevalence within the players with the Fermi rule linearly increases with the fraction of TFT players and explains the non-monotonic dependence in the structural mixing. Our study shows that the diversity in response behaviors may help to explain the emergence of cooperation in realistic contexts.

\end{abstract}




\end{frontmatter}


\section{Introduction}
Cooperation is central to the working of our societies and can be widely observed in biological, economic and social systems~\cite{Maynard1995The}. Deciphering its emergence and maintenance is a fundamental scientific question, which has attracted many researchers from different fields and now becomes a highly interdisciplinary field~\cite{Maynard1982Evolution,Gintis2000Game,rand2013human,perc2017statistical}. The key question to be addressed is: why do individuals help each other who could potentially be in competition and incur a cost to themselves?

Important progresses have been made with the help of evolutionary game theory~\cite{Nowak2004Evolutionary} by analyzing the stylized social dilemmas such as the prisoner's dilemma and the public goods game. Several mechanisms are proposed~\cite{Nowak2006Five} in the past several decades, such as reward and punishment~\cite{Sigmund2001Reward}, social diversity~\cite{Santos2008Social}, direct~\cite{trivers1971evolution} or indirect reciprocity~\cite{nowak1998evolution}, kin~\cite{hamilton1964genetical} or group selection \cite{keller1999levels,Queller1964Group}, spatial or network reciprocity~\cite{nowak1992evolutionary}. In particular, theoretically accounting for the fact that human populations are highly organized and individuals interact repeatedly with their immediate neighbors can support cooperation~\cite{nowak1992evolutionary}.  The rationale behind this is that a structured neighborhood facilitates the formation of cooperator clusters, which are able to effectively resist the invasion of defectors, as opposed to the well-mixed scenario. The ensuing years have witnessed a wealth of theoretical studies that further confirm this so-called network reciprocity for various population structures, including the multilayer networks~\cite{szabo2007evolutionary,Wang2015evolutionary}. Recently, the dynamical reciprocity as the counterpart mechanism is also proposed that points out that the interaction among co-evolving games could potentially lift the cooperation preference~\cite{liang2022dynamical}. 
It is worthy to note that recent human behavioral experiments give inconsistent results regarding the network reciprocity, structured populations do not promote cooperation in general~\cite{Gracial2012Human,Carlos2012Heterogeneous}, some additional conditions are required for cooperation to survive~\cite{Rand2014Static}. 

This unsatisfactory situation implies that some important elements could be missing in current game-theoretic models of realistic scenarios.
One element we would like to discuss in this work is the behavioral multimodality in response to the surroundings.  We use one behavior mode at one occasion or facing a person, but we could switch to another one for a different motivation.
For example, Ref.~\cite{Traulsen2010Human} shows that at the very beginning of their behavioral experiment, $62\%$ strategy updating can be explained by the rule of ``Imitate the best", the rest however are unexplainable by this, and the unexplained proportion increases around $4\%$ per round as the experiment goes on.
This confirms that our humans may use mixed modes of updating strategies rather than a single mode as assumed. 
All these observations suggest that the decision-makings in the real world could base upon a mixture of different modes rather than a single one assumed throughout the evolution in most previous theoretical models.

In fact, some recent theoretical works have noticed that the mode diversity in terms of strategy updating rules, and the mixed modes are investigated to see their impact on the emergence of cooperation. One example is the mixed modes of payoff-based imitation and conformity-based evolution~\cite{Szolnoki2015Conformity,Szolnoki2018Competition}, such mode mixture is potentially beneficial for the resolution of social dilemmas.
Yet, different impacts are seen when the modes of imitation and innovation are mixed, leading to the downfall of cooperation~\cite{Amaral2018Heterogeneous} or a cyclic dominance~\cite{Danku2018Imitate}. 
Another example is to investigate the population in the mixture of normal mode with ``irrational" one, where players act as zealots~\cite{Masuda2012evolution} or ``good Samarians"~\cite{Zheng2022probabilistic,Zheng2022pinning}, they show that the individuals in the irrational mode bring disproportionate promotion on the evolution of cooperation or fairness.
The general question to be addressed is: what is the impact of behavioral multimodality on the evolution of cooperation, what new complexities the mode mixing will bring?

In our work, we focus on the role of behavioral multimodality in the evolution of cooperation. Specifically, we consider a population with a bimodal mixture of Fermi~\cite{szabo1998evolutionary,szabo2005phase} and Tit-for-tat (TFT) updating rules~\cite{Axelrod1980Effective,Axelrod1980More,Axelrod1981The}. While the Fermi rule represents imitation learning mode, at the heart of many actions, players within the mode of TFT rule just repeat what their neighbors did to them in the previous round, both widely adopted in game-theoretic models. In the structural mode mixture, each player is endowed with a fixed rule, either Fermi or TFT rule; alternative, the two modes are probabilistically chosen by each player at each step. We find that in both implementations, the cooperation prevalence is considerably promoted compared to the single-mode scenario, especially, full cooperation is obtained in a wide range of parameters in the latter case. We also develop a mean-field treatment that correctly reproduces the observations for the former case. 


\section{Bimodal response behavior model}
We study the prisoner's dilemma (PD) game with $N$ individuals that are located on an $L \times L$ square lattice with a periodic boundary condition. PD is a typical pairwise game for many social dilemmas, mutual cooperation brings the reward $R$, mutual defection yields the punishment $P$ , and mixed encounter gives the cooperator the sucker's payoff $S$ yet the temptation $T$ for the defector.  $T>R>P>S$ and $2R>T+S$ are required for PD. It's easy to find that defection as the Nash equilibrium is the better choice regardless of the opponent's selection yet the mutual cooperation is optimal for their collective profit. We adopt the weak version of PD, $R=1$ , $P=S=0$, and $T=b$, where $1.0 \leq b \leq 2.0$ ~\cite{nowak1992evolutionary}. 

In the previous practice, each individual is assigned a strategy, either cooperation (C) or defection (D) opposing to all its neighbors, which is the \emph{node-based} strategy with the strategy set $\mathbb{S}_n=\{C,D\}$. Here, we extend this setup to the \emph{edge-based} strategy that a player can use different strategies against different neighbors (i.e., along different edges), which is more commonly seen in the real world. In the edge-based setup, the state of an individual $i$ is characterized by the fraction of cooperation strategy against its neighbors defined as $s_{i}={n_i(C)}/{k_i}$, where $n_i(C)$ is the number of edge strategies for the player $i$ with C and $k_i=4$ on the $2d$ square lattice. Thus, $s_{i}\in \mathbb{S}_e=\{0,0.25,0.5,0.75,1\}$, which can be interpreted as \emph{the cooperation propensity}, $s_i=1$ and $0$ correspond to the C and D strategies respectively in the node-based strategy set $\mathbb{S}_n$. 

For simplicity, we build a bimodal behavior model by mixing Fermi and TFT rules, which are two commonly used responses. Specifically, the mixture is implemented in two ways --- structural mixing (SM) and probabilistically mixing (PM). In the SM implementation, each player randomly chooses TFT rule as their acting mode with a probability $\omega$, and with the Fermi rule otherwise. They stay in the same mode throughout the evolution. By contrast, in the PM implementation, players are all identical, they probabilistically adopt TFT and Fermi rules respectively with the probability $\omega$ and $1-\omega$ in every single step.    

At the very start, each edge strategy of every player is randomly assigned with C or D towards their neighbors with an equal chance. An elementary step of the Monte Carlo simulation for the SM implementation is as follows. A player $i$ is randomly chosen, if player $i$ is with TFT rule, player $i$ will adopt the edge strategies that all its neighbor plays against it, either C or D, according to the TFT rule; i.e., $s_{i}={n'_i(C)}/{k_i}$, where $n'_i(C)$ is the number of edge strategy with C that its neighbors playing against to $i$. 
Otherwise, the evolution of player $i$'s strategy is based upon the Fermi rule as follows.
First, one of $i$'s neighbors $j$ is randomly selected, player $i$ and $j$ respectively acquire their mean payoff $\bar{\Pi}_{i,j}$ (defined by their total payoffs $\Pi_{i,j}$ divided by their degrees). Next, player $i$ adopts the cooperation propensity of player $j$ with the probability ~\cite{szabo1998evolutionary,szabo2005phase}
\begin{equation}\label{eq:imitation}
	W(s_{j}\rightarrow s_{i})=\dfrac{1}{1+\exp[(\bar{\Pi}_{i}-\bar{\Pi}_{j})/K]},
\end{equation}
where $K$ is a temperature-like parameter, which can be interpreted as the environment uncertainties in the imitation process, and will be fixed at $0.025$ throughout the study. Lastly, the strategies against the four neighbors of player $i$ are pinned down according to the newly adopted propensity $s_j$ if player $j$'s strategy is successfully imitated, i.e., each of its edge strategy chooses the strategy C according to the probability $s_j$ independently. None of $i$'s strategy will be updated if the imitation is unsuccessful. Notice that, the imitation in the edge-based Fermi rule is to copy the cooperation propensity not the four strategies, which means the resulting $s_i$ could be unequal to $s_j$, unless the propensity is 0 or 1.    

The procedure for PM differs only at the beginning stage of every elementary step. The randomly chosen player $i$ evolves according to the TFT rule with a probability $\omega$, and with $1-\omega$ updates according to the edge-based Fermi rule. For the TFT case, it copies exactly what their neighbors' edge strategies towards it; For the edge-based Fermi rule, it imitates the cooperation propensity of one random neighbor $j$ with the same way in the above SM.

Note that, the above model simulation follows a typical asynchronous updating procedure. A complete Monte Carlo step (MCS) consists of $N$ elementary steps, meaning that every player updates its state exactly once on average. We compute the cooperation prevalence $f_C=\frac{1}{N}\sum_{i=1}^{N} s_i$ as the primary order parameter, measuring the overall preference in cooperation of the population. The total sampling time is $10000$ MCSs and the equilibrium density of cooperation is obtained by averaging over the last $1000$ MCSs.

\section{Results}

\subsection{2d square lattice}
\begin{figure}[!h]
	\centering
	\includegraphics[width=0.32\linewidth]{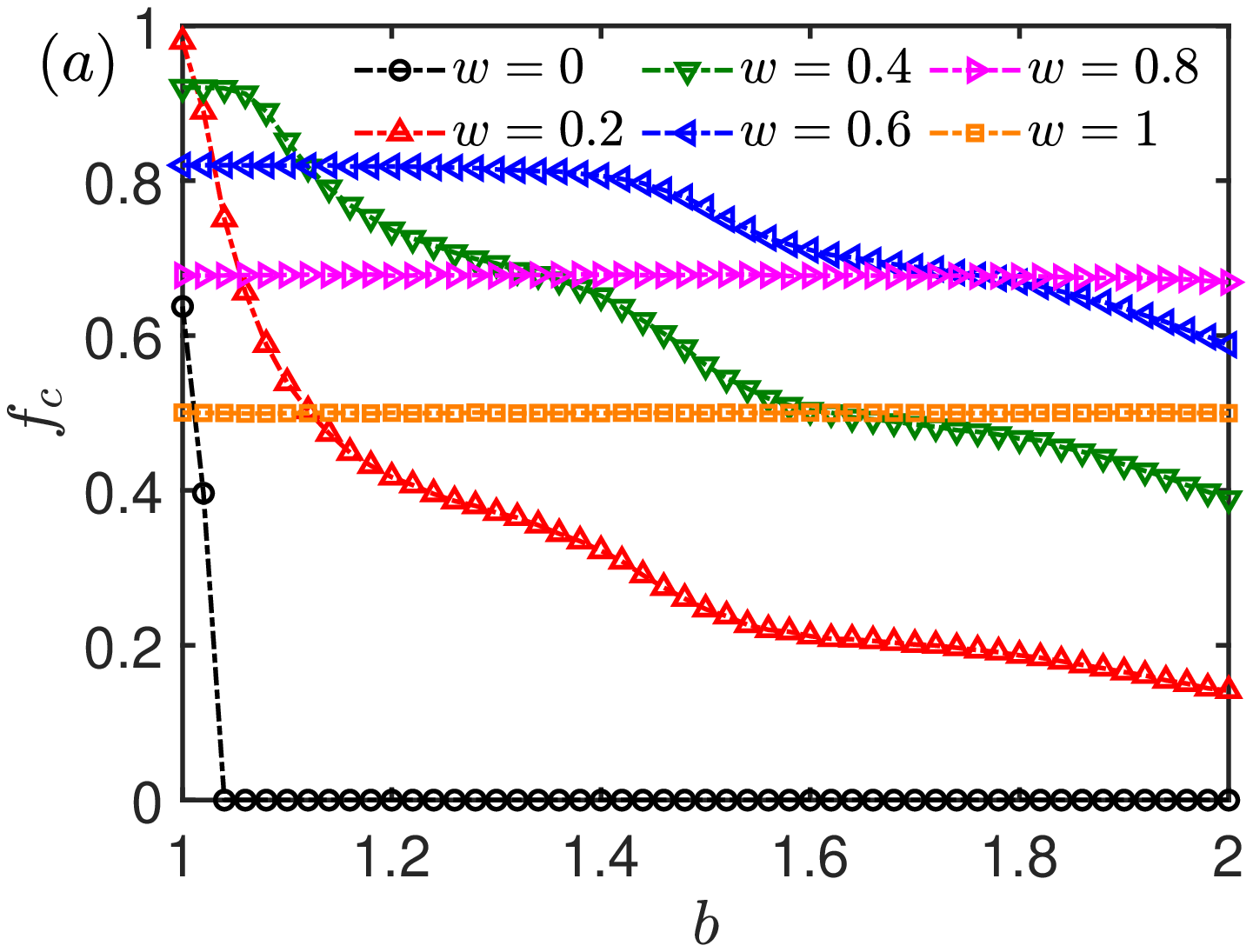}
	\includegraphics[width=0.32\linewidth]{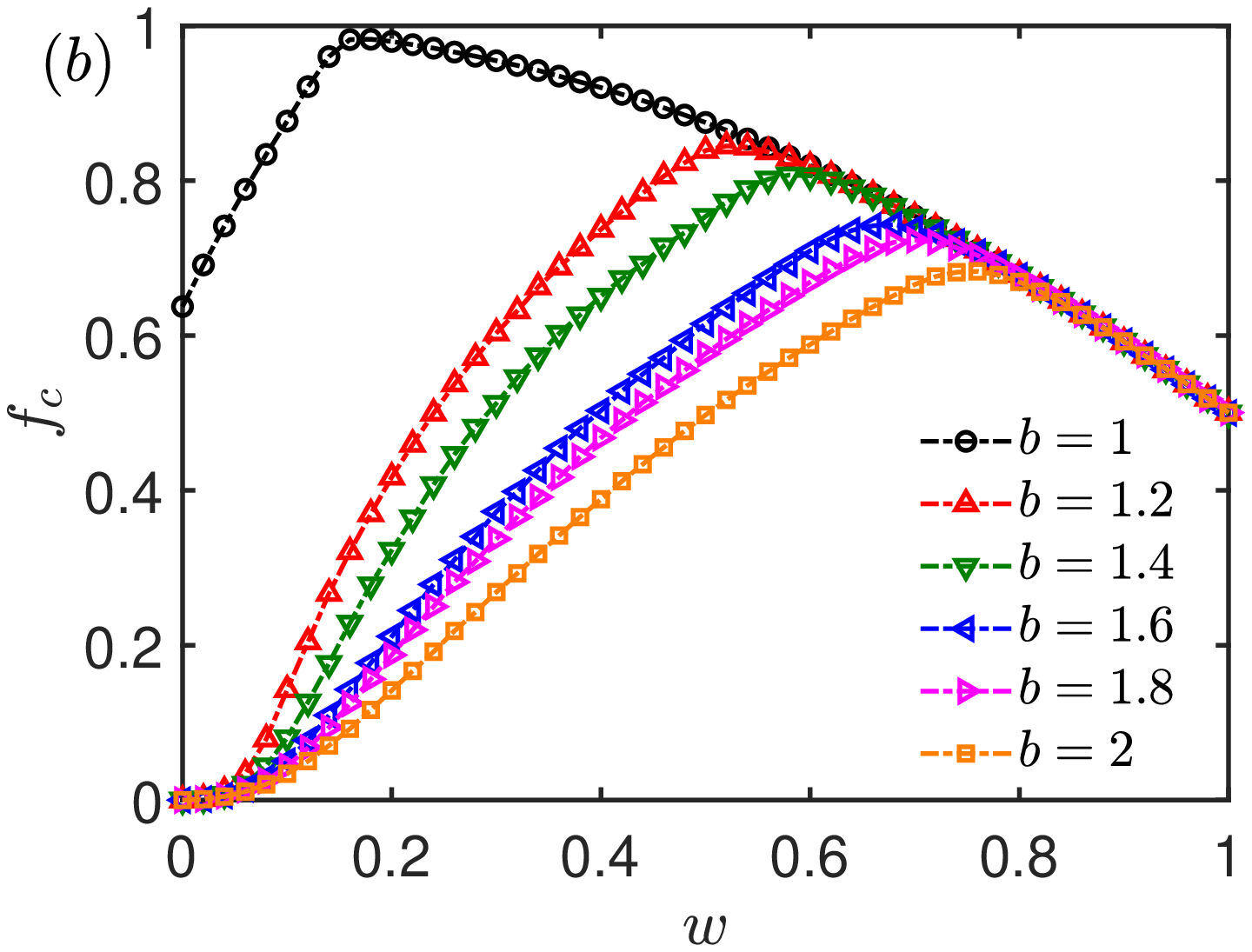}
	\includegraphics[width=0.32\linewidth]{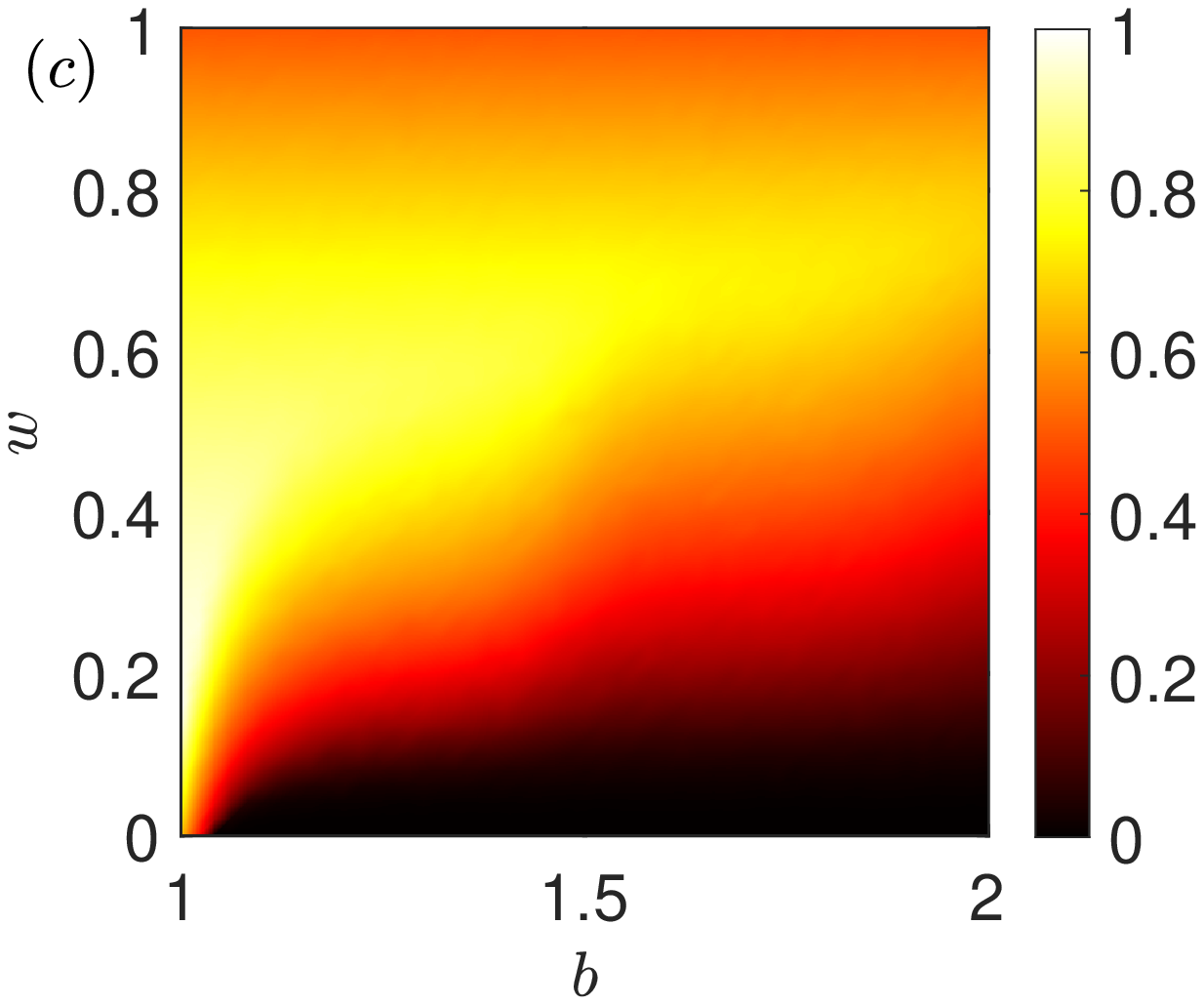}
	\caption{The evolution of cooperation on $2d$ square lattice with SM. 
		(a) The prevalence of cooperation $f_c$ as a function of temptation $b$;
		(b) $f_c$ as a function of $\omega$, the fraction of TFT players; 
		(c) Heat map for the cooperation prevalence $f_c$ in the $b-\omega$ parameter space. 
		Other parameters: $L=1024$ for (a,b) and 256 for (c).
	}
	\label{Fig:JGpfp}
\end{figure}

\emph{Structural mixing (SM) --} Fig.~\ref{Fig:JGpfp} reports the results on the 2d square lattice in the case of structural mixing. Let's first see the two extreme cases ($\omega=0,1$), the single-mode scenarios. As can be seen in Fig.~\ref{Fig:JGpfp}(a), when the population all act according to the Fermi updating rule ($\omega=0$), cooperation can only survive in a narrow region $b<b_c\approx1.037$, which is exactly the same as previous studies using node-based Fermi rule~\cite{szabo2005phase} 
(
see Sec. I in Supplemental Material). This means that the replacement of node-based strategy with the edge-based version per se does not bring any change in $f_c$.  When all individuals use the mode of the TFT rule ($\omega=1$), the prevalence of cooperation $f_c$ is always approximately $0.5$ regardless of the value of temptation $b$, because the initial level of cooperation $f_c(t=0)\approx 0.5$ is basically reserved according to the TFT rule. As the two modes are mixed $0<\omega<1$, the cooperation levels are all lifted than the pure Fermi rule case, surprisingly there are some mixtures (e.g. $\omega=0.6$, 0.8) that can leads to $f_c>0.5$, higher than the expected level in the pure TFT case. Fig.~\ref{Fig:JGpfp}(b) show explicitly  that there exists optimal mixing ratio $\omega_{o}$ that leads to the highest level of cooperation $f_c$. As expected, $f_c$ is reduced as the temptation $b$ is increased, whereas the value of $\omega_{o}$ is shifted to be larger.
The dependence of $f_c$ on the two parameters are summarized in the phase diagram shown in Fig.~\ref{Fig:JGpfp}(c).

\begin{figure}[!h]
	\centering
	\includegraphics[width=0.45\linewidth]{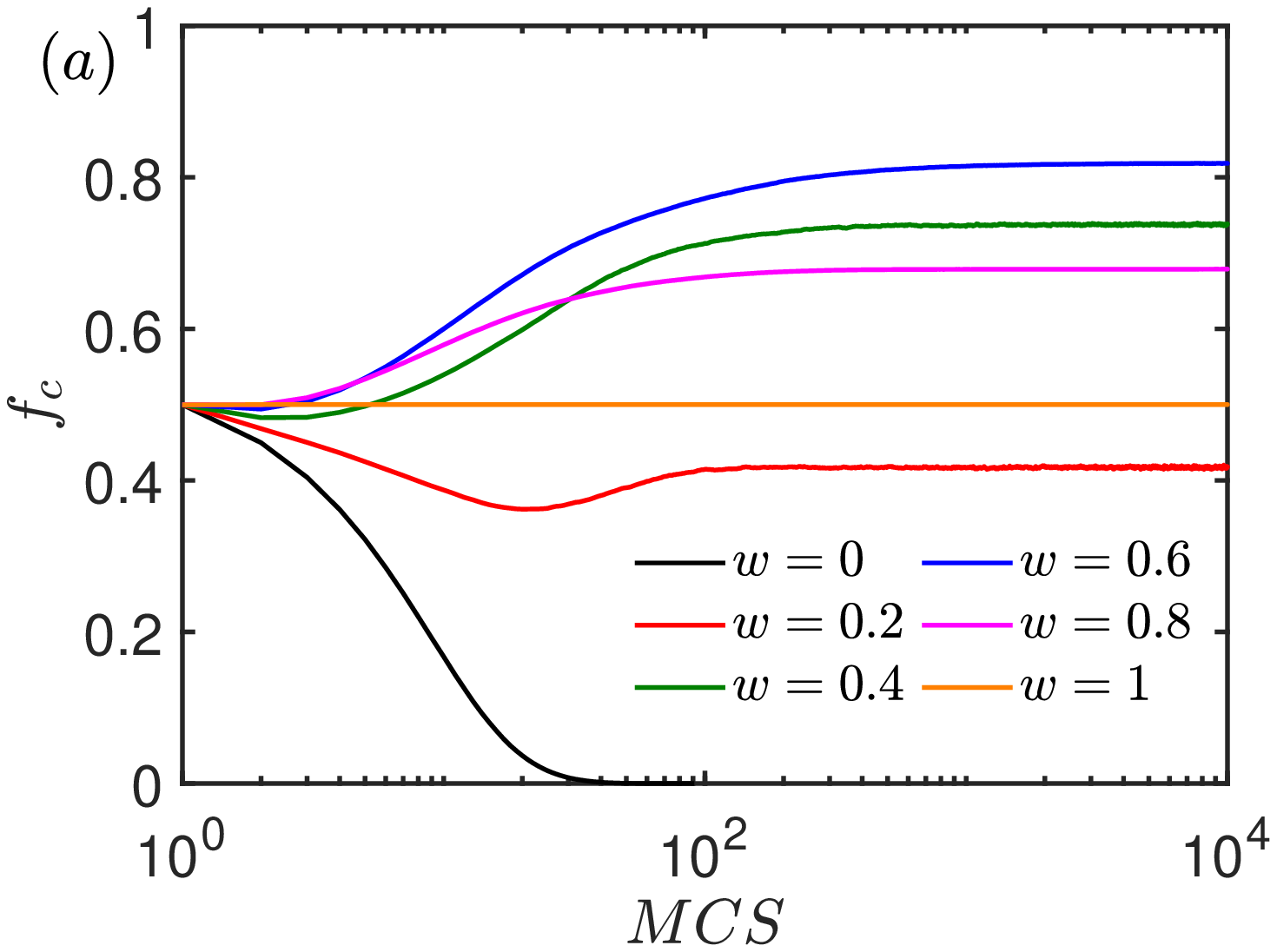}
	\includegraphics[width=0.45\linewidth]{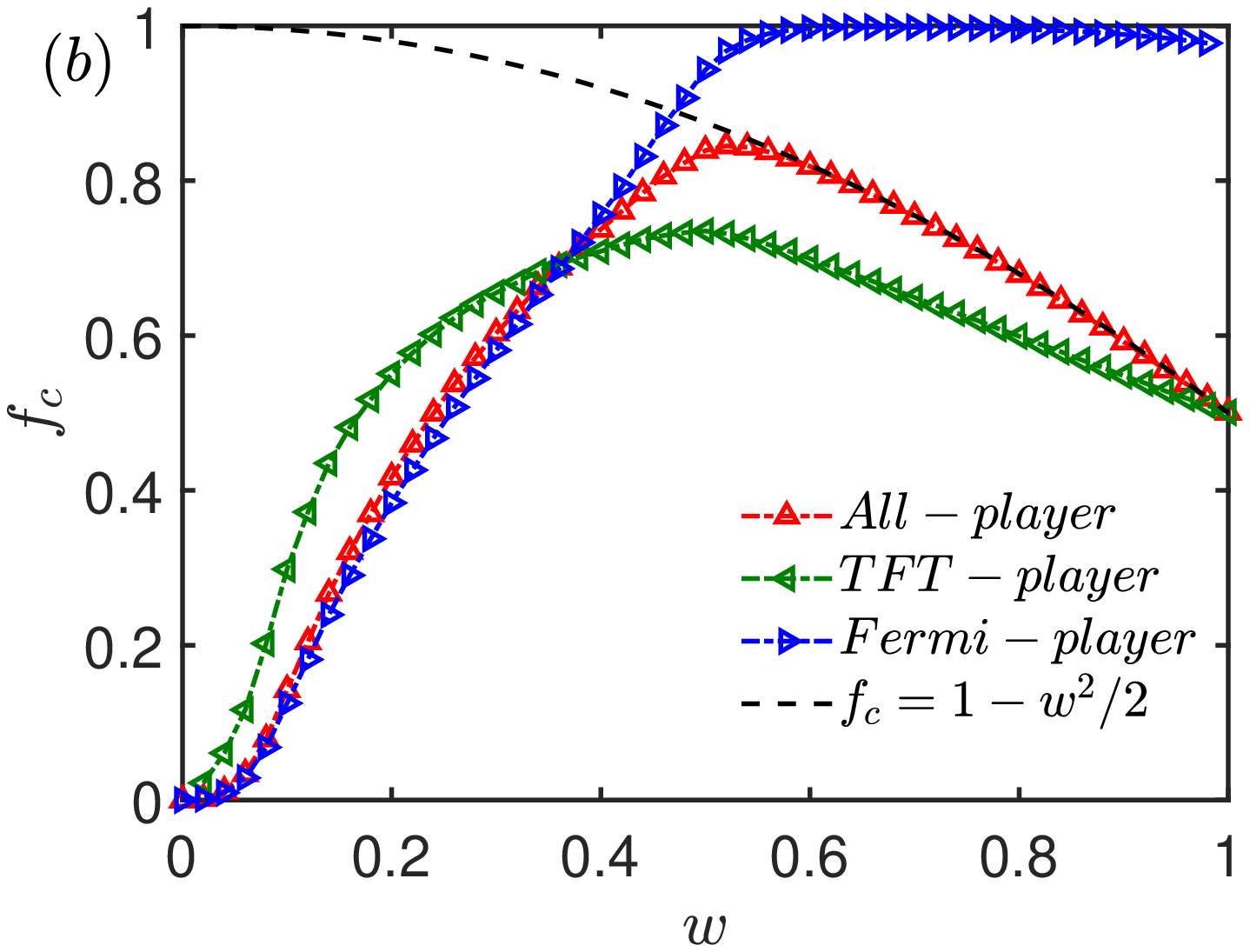}
	\caption{Further analysis of the cooperation evolution on the $2d$ square lattice with SM.
		(a) Time series; (b) The prevalence of cooperation $f_c$ versus $\omega$ computed separately for TFT and Fermi players, together with the whole population for comparison. The black dashed line is the fitting line.
		Parameters: $L=1024$, $b=1.2$.
	}
	\label{Fig:JGtf}
\end{figure}

The non-monotonic dependence of cooperation prevalence on $\omega$ is confirmed by the typical time series by fixing $b=1.2$, see Fig.~\ref{Fig:JGtf}(a). For better understanding, the cooperation prevalence for the two modes is respectively shown in Fig.~\ref{Fig:JGtf}(b).
As can be seen, when $\omega<\omega_{o}$ ($\omega_{o}\approx 0.52$ in this case), the values of $f_c$ for both modes increase. But once $\omega>\omega_{o}$, the players with the Fermi rule are in almost full cooperation state (i.e., $s_i\approx 1$), whereas the cooperation prevalence of TFT players decreases from the peak value to the expected level 0.5. Once $s_i=1$ for Fermi players, the strategies between Fermi and TFT players are also all cooperation, the defection  then only comes from the TFT players. By estimation, about half TFT-TFT interaction edges finally will be in the deadlock of mutual defection D-D state, while the other half in C-C state. Therefore, 
\begin{equation}\label{eq:fc_approx_right}
	\begin{aligned}
		f_c\approx 1-\omega^2/2,
	\end{aligned}
\end{equation}
the approximation for the overall cooperation prevalence when $\omega>\omega_{o}$, which fits very well with the numerical results in Fig.~\ref{Fig:JGtf}(b).

To develop the intuition of why mixing promotes cooperation, some typical spatial patterns are provided in Fig.~\ref{Fig:JGpattern}, where the states for all players together with TFT- and Fermi-players are respectively shown in different columns. A critical observation is the difference in cooperation prevalence between TFT- and Fermi-players. By combination, all interactions within the mixing populations can be classified into three types: 

\begin{figure}[!h]
	\centering
	\includegraphics[width=0.9\linewidth]{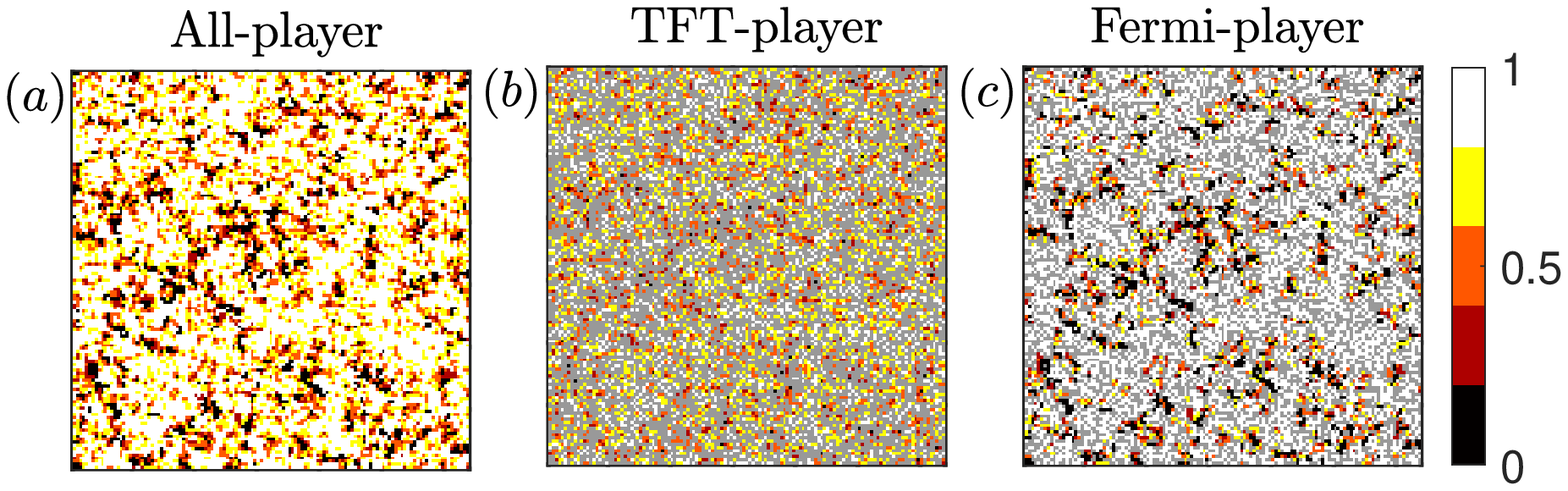}\\
	\includegraphics[width=0.9\linewidth]{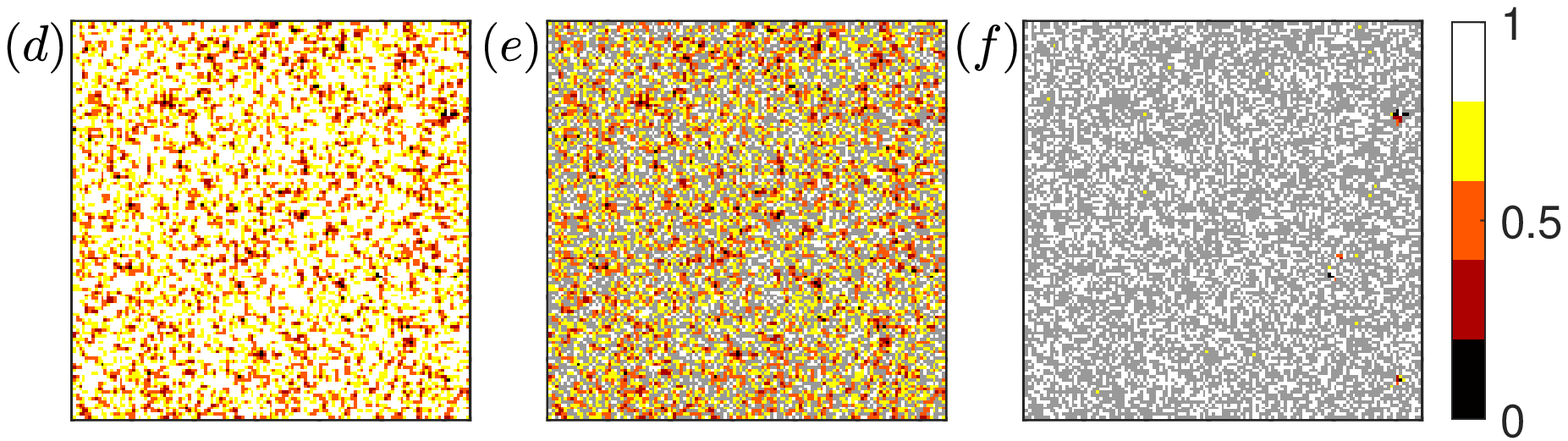}\\
	\includegraphics[width=0.9\linewidth]{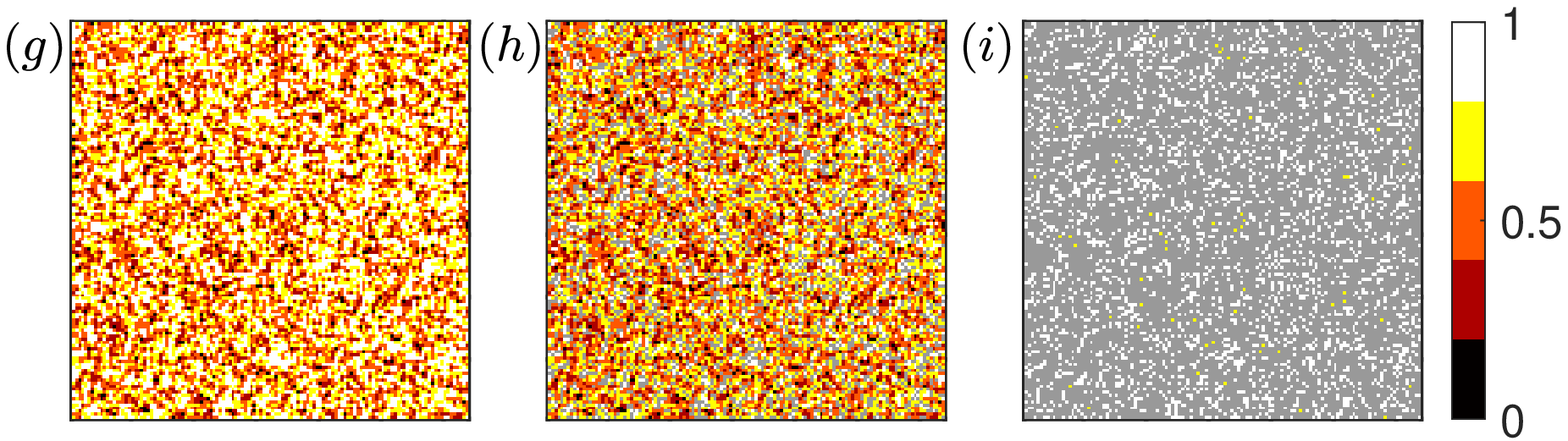}
	\caption{The spatial patterns on $2d$ square lattice with SM for different $\omega$.
		(a-c) $w=0.4$, (d-f) $w=0.6$, (g-i) $w=0.8$. 
		The first column is the states of all players, the colors (black, crimson, orange, yellow, white) represent the cooperation propensity of players $s_i$ ($0,0.25,0.5,0.75,1$). 
		The second column is only for TFT-players, where the gray sites are players using the Fermi rule. 
		The third column is only for Fermi-players, also the gray sites are those using the TFT rule. 
		Other parameters: $N=128 \times 128$, $b=1.2$.
	}
	\label{Fig:JGpattern}
\end{figure}

i) \emph{Fermi-Fermi interactions}: when only Fermi-players are present, they compute, compare and imitate, finally the population evolve into the Nash equilibrium point, the full defection solution, only the network reciprocity may help (but not the case for $b=1.2$ here). 

ii) \emph{TFT-TFT interactions}: since TFT players just repeat what their opponents have done to them, the edge strategies will be frozen as either C-C or D-D pairs in the asynchronous updating, irrespective of the parameter $b$. The value of $f_c\approx f_c(t=0)$ for random initial conditions, and $f_c(t=0)\approx0.5$ in our study.   

iii) \emph{Fermi-TFT interactions}: due to the random initialization, TFT-players have diverse payoffs, and those of high cooperation propensity $s_i$ generally have higher payoffs than those of small $s_i$, which are more likely to be imitated by their Fermi-neighbors (see e.g. white sites in Fig.~\ref{Fig:JGpattern}(c)). This in turn increases their neighbors' payoff and improve their values in $s_i$ through both type i) and ii) interactions, and drives the overall cooperation prevalence to rise in the end. 

This means that if the fraction of Fermi-players is too large (i.e., $\omega\rightarrow 0$), Fermi-Fermi interactions deteriorate the cooperation; likewise, if TFT-players are too much ($\omega\rightarrow 1$), $f_c\rightarrow 0.5$ due to the outcome of TFT-TFT interactions. Only when the two fractions are comparable, Fermi-TFT interactions then come into play that improve cooperation; especially $\omega\approx \omega_o$,  a balance point is reached to reduce both type i) and ii) interactions effectively. 
Otherwise, when $\omega<\omega_o$, Fermi-players still form sizeable clusters that Fermi-Fermi interactions deteriorate the cooperation (Fig.~\ref{Fig:JGpattern}(a)). 
On the contrary, when $\omega>\omega_o$, even though all Fermi-TFT edge strategies are still largely within the C-C pairs (Fig.~\ref{Fig:JGpattern}(i)), more TFT-TFT interactions yield more defection (Fig.~\ref{Fig:JGpattern}(h)), reducing the overall cooperation. 
This explains the non-monotonic dependence of $f_c$ on the mixing ratio $\omega$ and the existence of optimal value $\omega_o$, shown in Fig.~\ref{Fig:JGpfp}(b).

\begin{figure}[!h]
	\centering
	\includegraphics[width=0.32\linewidth]{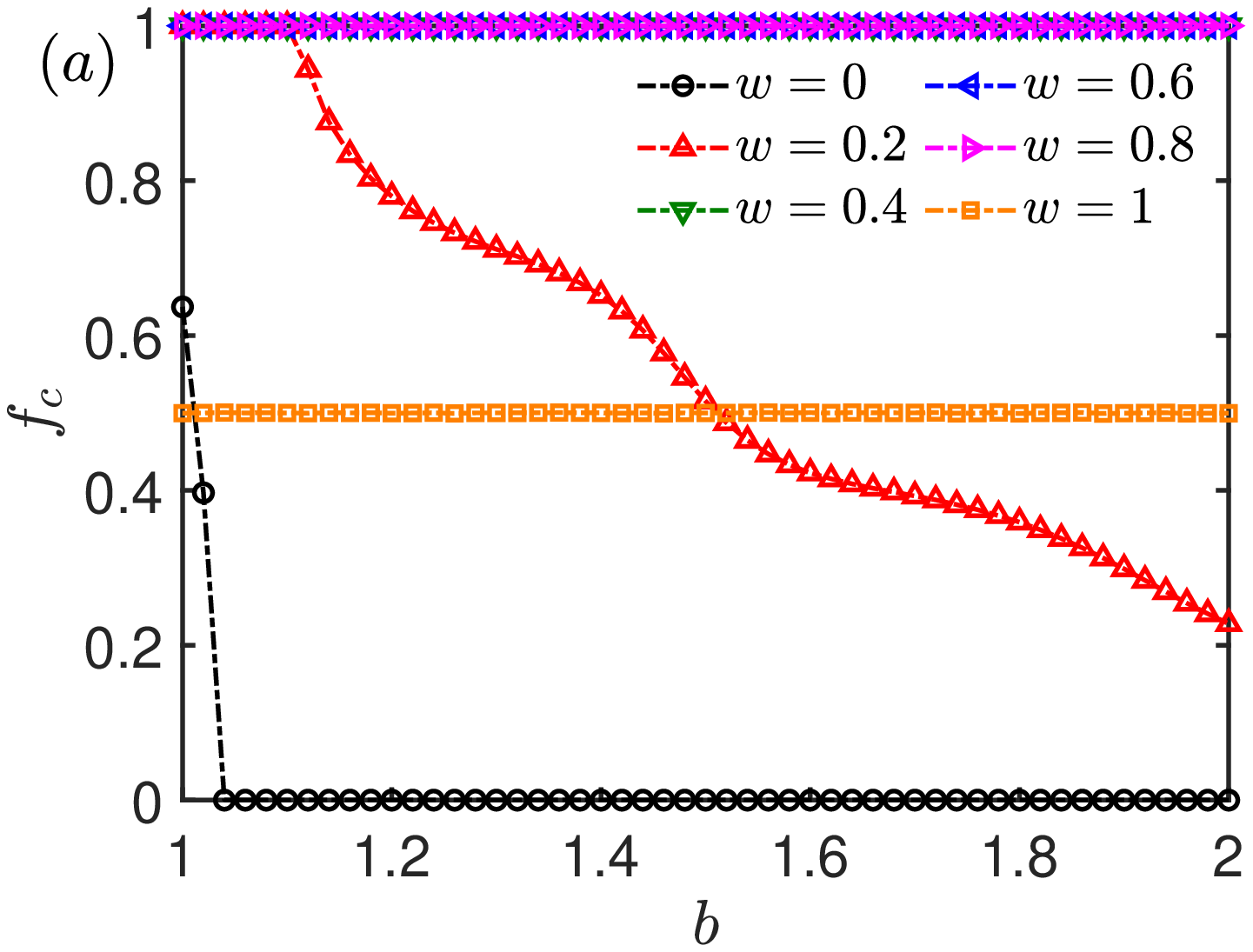}
	\includegraphics[width=0.32\linewidth]{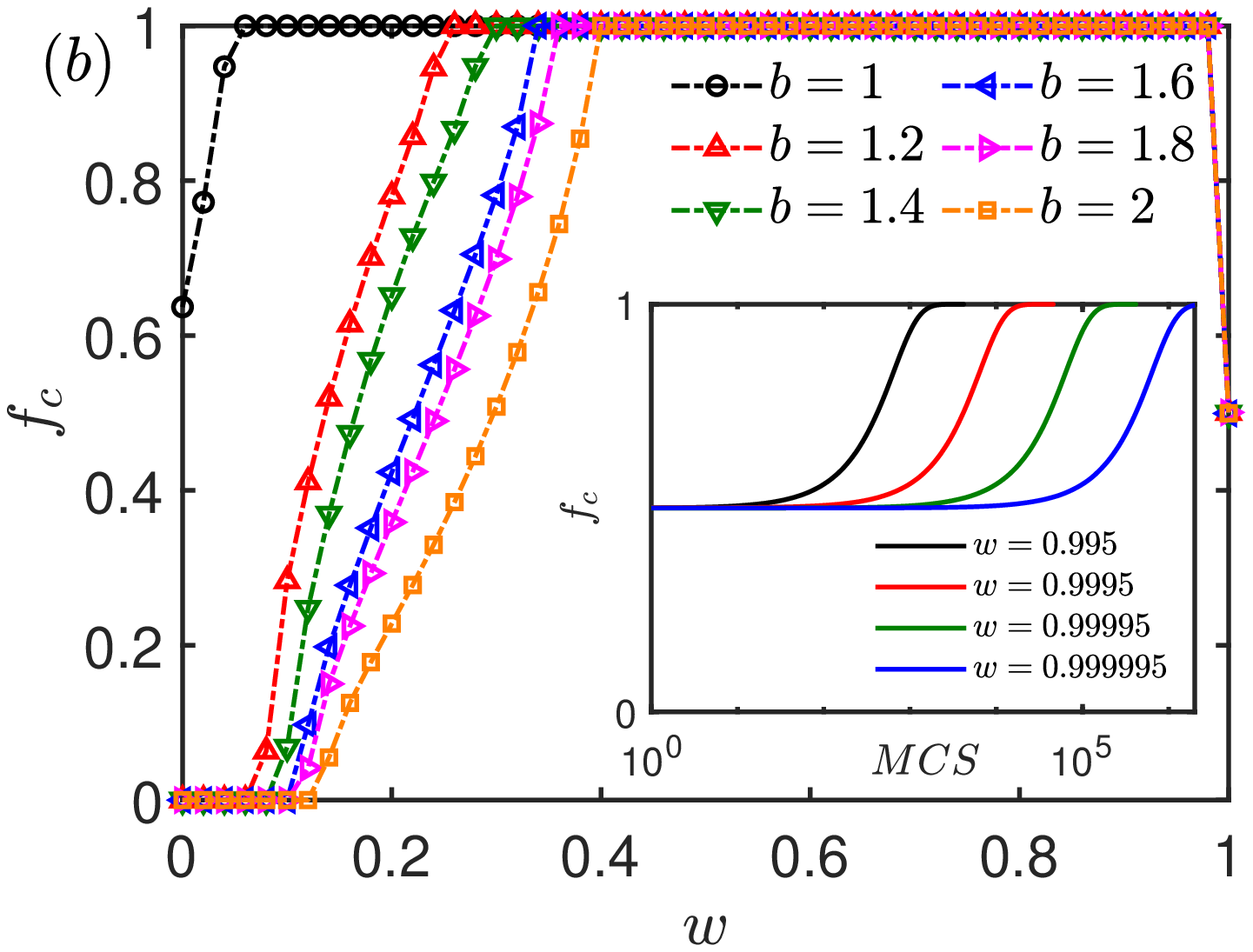}
	\includegraphics[width=0.32\linewidth]{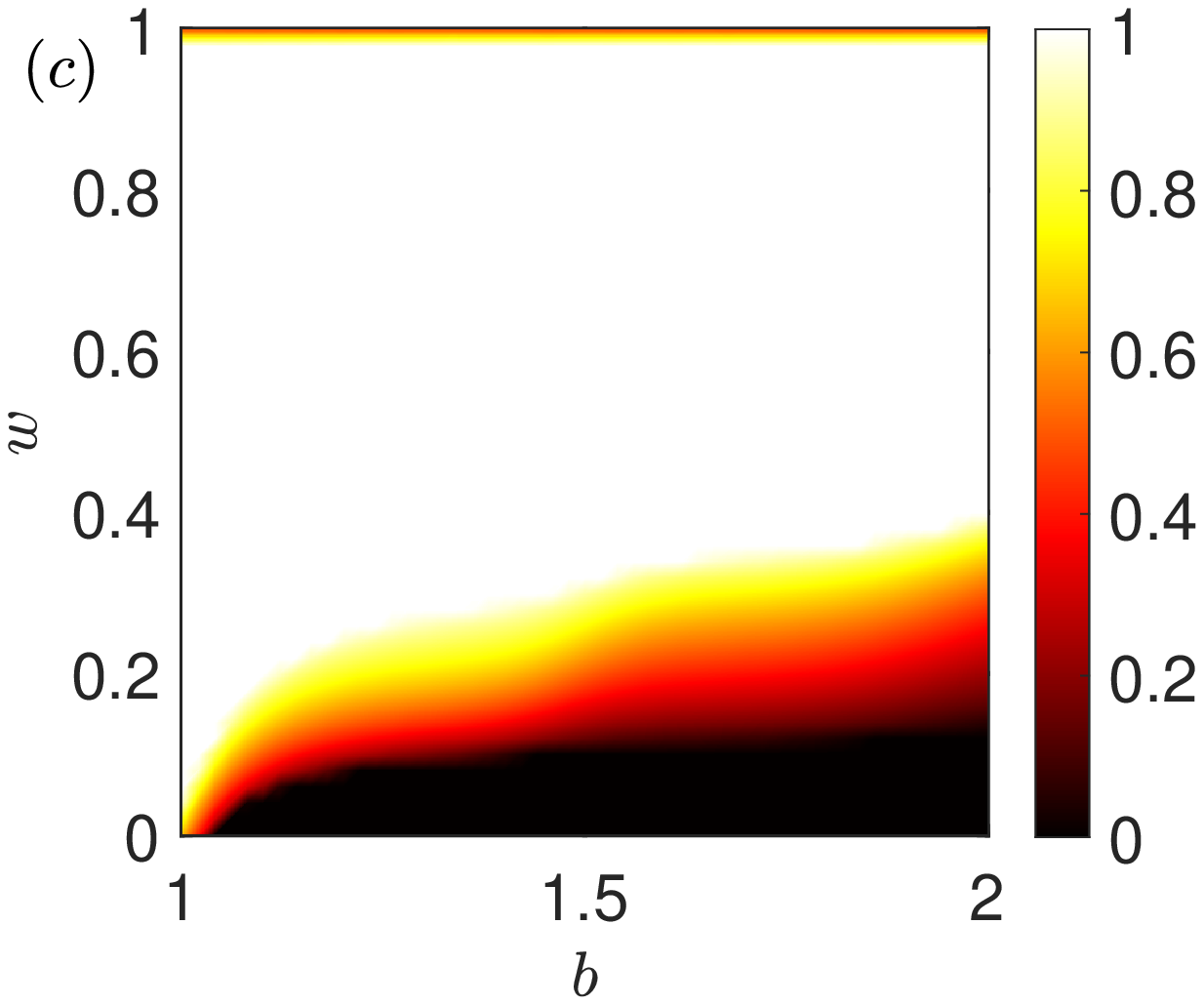}
	\caption{The evolution of cooperation on the $2d$ square lattice with PM.
		$\omega$ is the probability that individual evolves according to the TFT rule and to the Fermi rule otherwise at every single step.
		The prevalence of cooperation $f_c$ as a function of temptation $b$ (a) and $\omega$ (b); 
		The inset are several time series for $\omega$ being very close to 1, with fixed $b=1.2$.
		(c) Phase diagram for $f_c$ in $b-\omega$ parameter space. 
		Other parameters: $L=1024$ for (a,b) and 256 for (c).
	}
	\label{Fig:GLpfp}
\end{figure}

\emph{Probabilistic mixing (PM) -- } The results of probabilistic mixing are shown in Fig.~\ref{Fig:GLpfp}, where $\omega$ is now interpreted as the probability to behave within the mode of TFT rule at every single step. 
Fig.~\ref{Fig:GLpfp}(a) provides the cooperation prevalence $f_c$ as the function of the temptation $b$ for a couple of $\omega$, where $f_c$ increases and $f_c=1$ for $\omega=0.4,0.6,0.8$, irrespective of $b$. Fig.~\ref{Fig:GLpfp}(b) further shows the dependence of $f_c$ on $\omega$, beyond a critical value  $\omega_c$ an absorbing state of full cooperation is reached except $\omega=1$, which is recovered to the case of pure TFT-TFT interactions (type ii). To examine this extreme, several time series for $\omega$ being very close to 1 are shown in the inset by fixing $b=1.2$, which indicates that full cooperation can always be reached as long as $\omega<1$, just a long transient is needed when $\omega\rightarrow 1$. The reason for the absence of $\omega_o$ in the PM, is because now the listed three types of interactions above are no fixed anymore; therefore there is no permanent deadlock of D-D in TFT-TFT interactions, and the cooperation prevalence can sustain at $f_c=1$ once $\omega_c<\omega<1$. The full dependence of $f_c$ on these two parameters are summarized in the phase diagram Fig.~\ref{Fig:GLpfp}(c).

\begin{figure}[!h]
	\centering
	\includegraphics[width=0.6\linewidth]{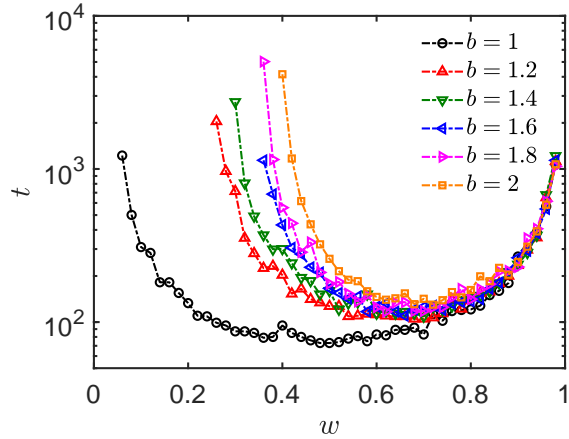}
	\caption{The converging time versus $\omega$.
		The time for the population to reach the full cooperation $f_c=1$ as a function of the probability $\omega$, on the $2d$ square lattice with PM.
		$L=1024$.
	}
	\label{Fig:GLtw}
\end{figure}

Although the absorbing state is always reached within the region $\omega_c<\omega<1$, the converging time still depends on $\omega$. Fig.~\ref{Fig:GLtw} gives the time needed when starting from random initial conditions, showing that there exists an optimal value of $\omega$, with which the transient is the shortest.  

\begin{figure*}[!h]
	\centering
	\includegraphics[width=0.45\linewidth]{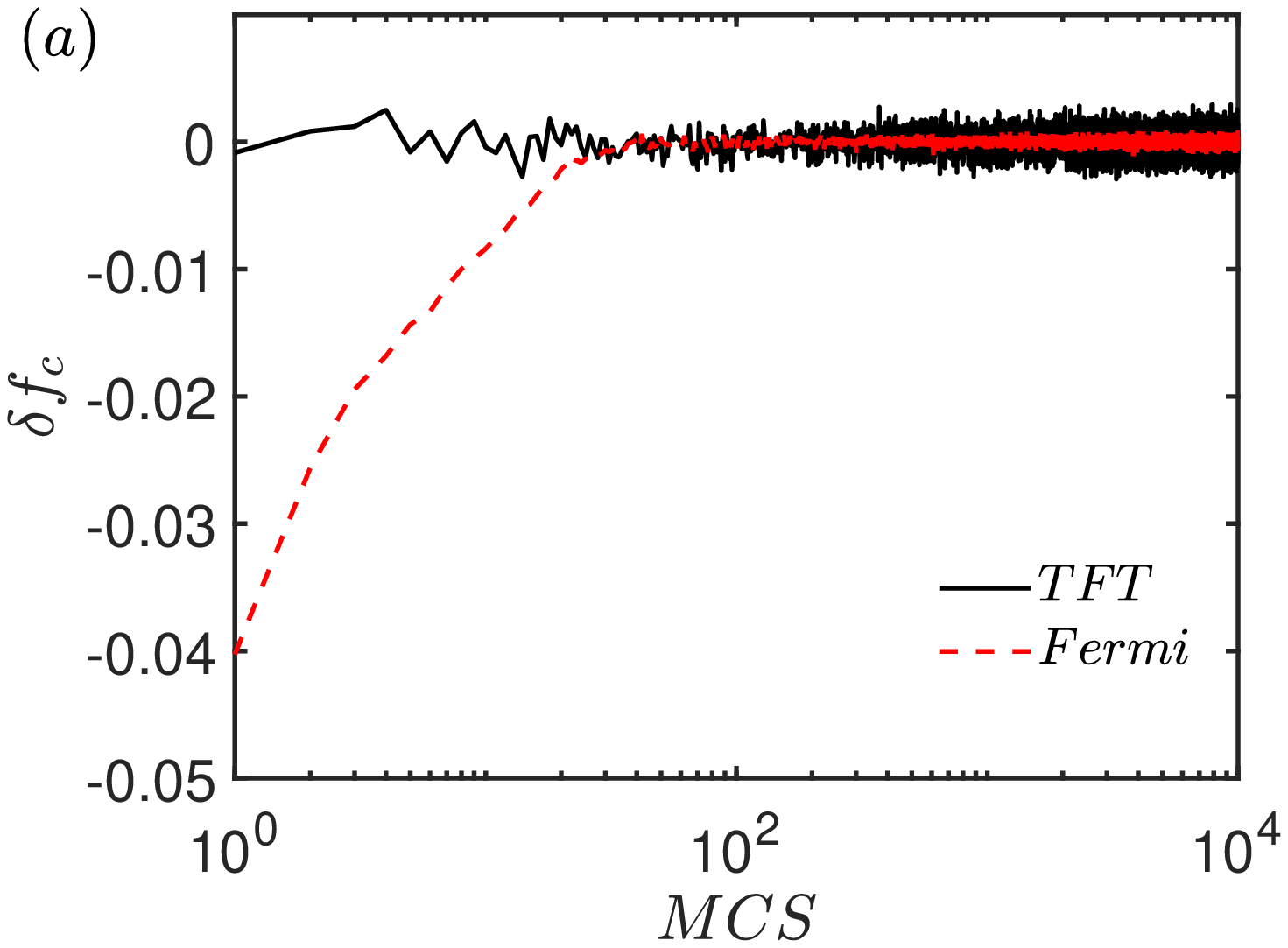}
	\includegraphics[width=0.45\linewidth]{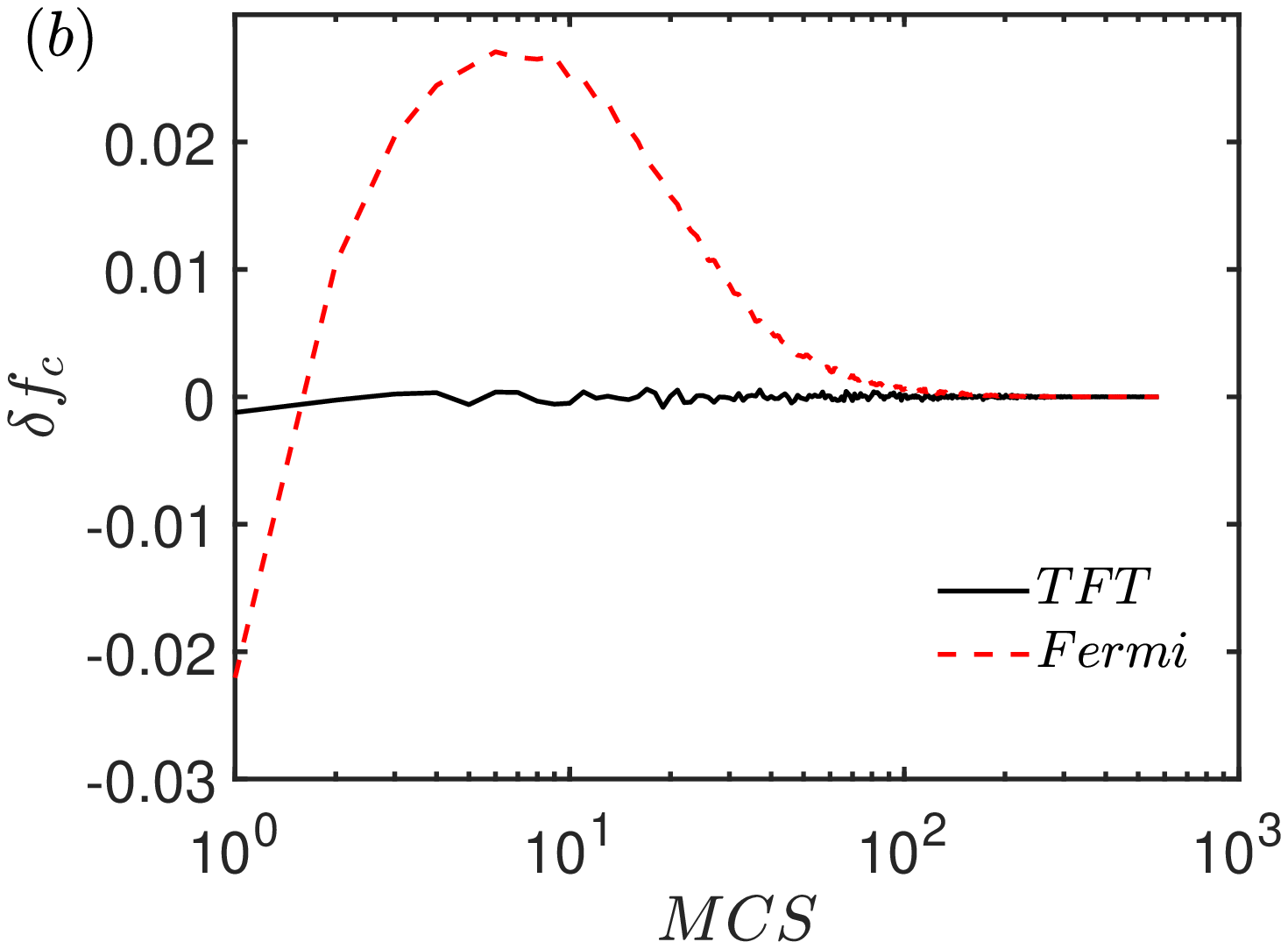}\\
	\includegraphics[width=0.45\linewidth]{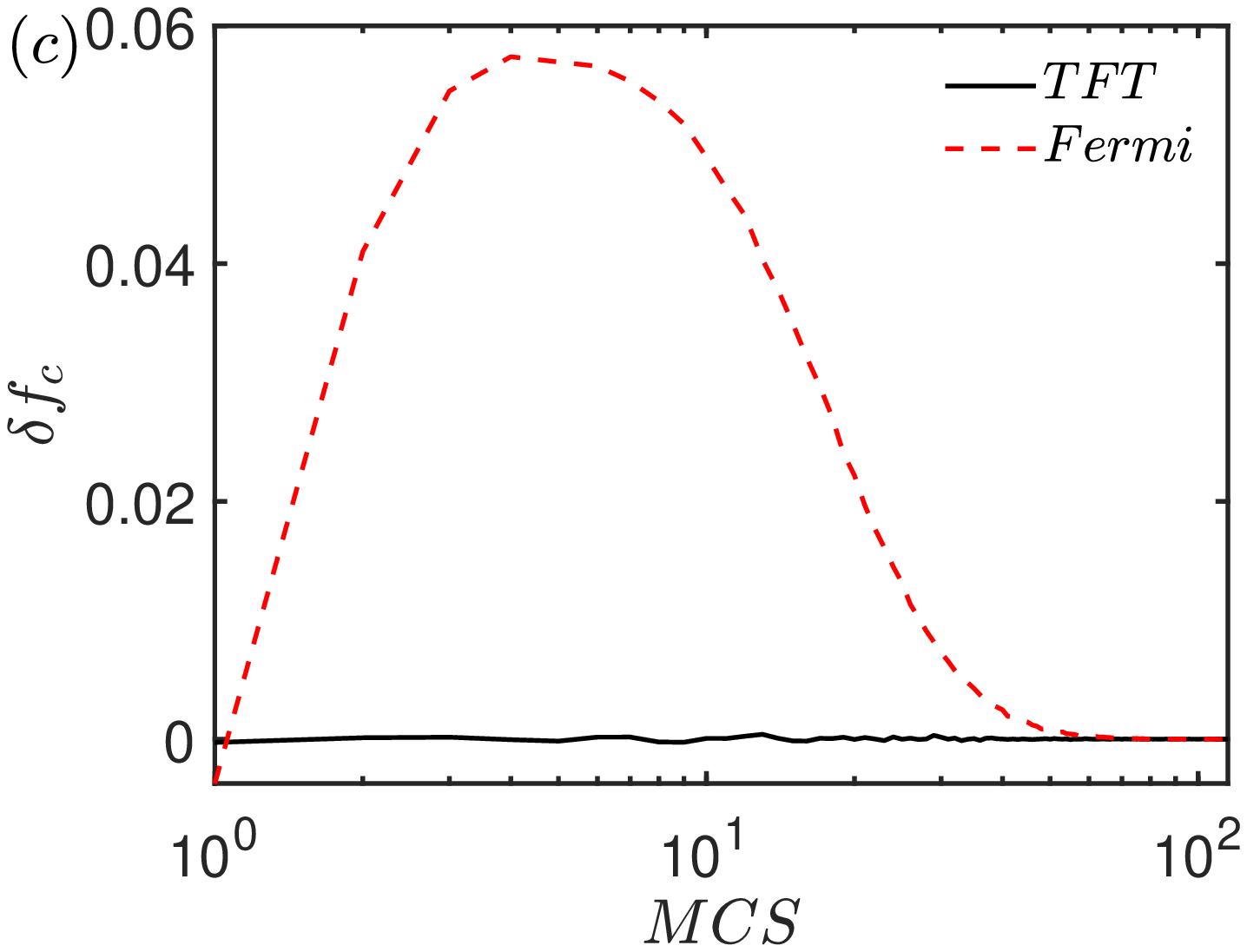}
	\includegraphics[width=0.45\linewidth]{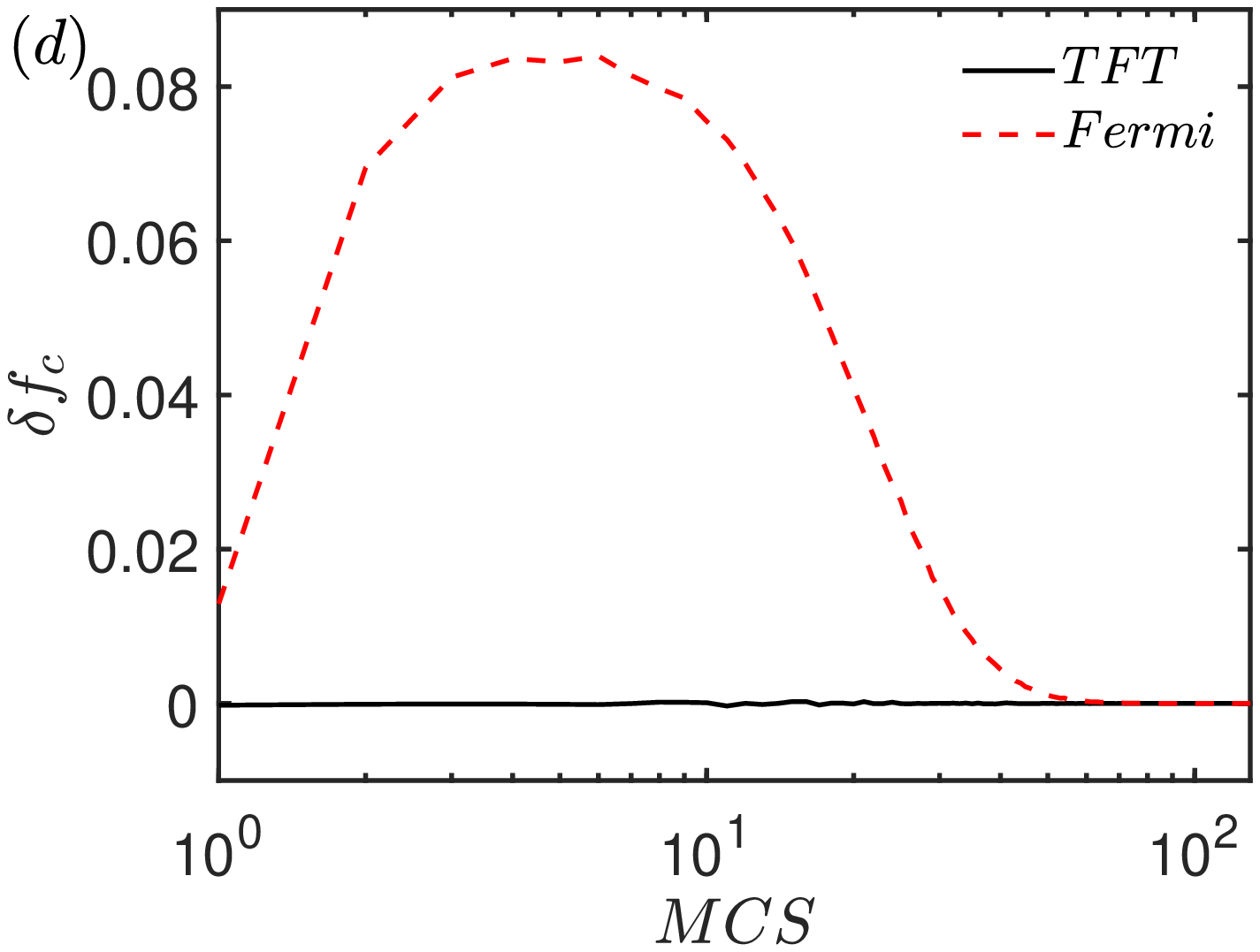}
	\caption{The contribution comparison in the cooperation evolution.
		The average value of $\delta f_c=\langle s_i(\text{new})-s_i(\text{old})\rangle$ per MCS on the 2d square lattice with PM for different $\omega$.
		(a) $w=0.1$, (b) $w=0.3$, (c) $w=0.5$, and (d) $w=0.7$. 
		Parameters: $L=1024$ and $b=1.2$.
	}
	\label{Fig:GLdeltafc}
\end{figure*}

To better understand the mechanism behind, we estimate the changes in the cooperation propensity per step, respectively for Fermi- and TFT-rule, shown in Fig.~\ref{Fig:GLdeltafc}. Specifically, $\delta f_c$ is the average change in $s_i$ for every single update over an MCS. It shows that the main contribution for the cooperation promotion comes from the actions based on the Fermi-rule, while the changes from TFT-rule are much less pronounced. When $\omega$ is small (see Fig.~\ref{Fig:GLdeltafc}(a)), $\delta f_c<0$, that the cooperation prevalence $f_c$ is low in this case; but if $\omega$ becomes larger, Fermi-players actively entrain the cooperation level to a pretty high level (Fig.~\ref{Fig:GLdeltafc}(b-d)).

\subsection{Complex networks}
To check the robustness of our observations, we also carry out numerical experiments on two complex networks. Specifically, we adopt Erd\H{o}s-R\'enyi (ER) random networks~\cite{erdos1960on} and Barab\'asi-Albert (BA) scale-free networks~\cite{barabasi1999emergence}, respectively represent homogeneous and heterogeneous networks in the real world. As is well-known, the degrees of ER networks satisfy poisson distribution, whereas the degree distribution of BA networks is a power-law with an exponent of $-3$. For the ease of comparison, both network sizes are $N=2^{20}$ with the same average degree $\langle k\rangle=4$. 

\begin{figure*}[!h]
	\centering
	\includegraphics[width=0.45\linewidth]{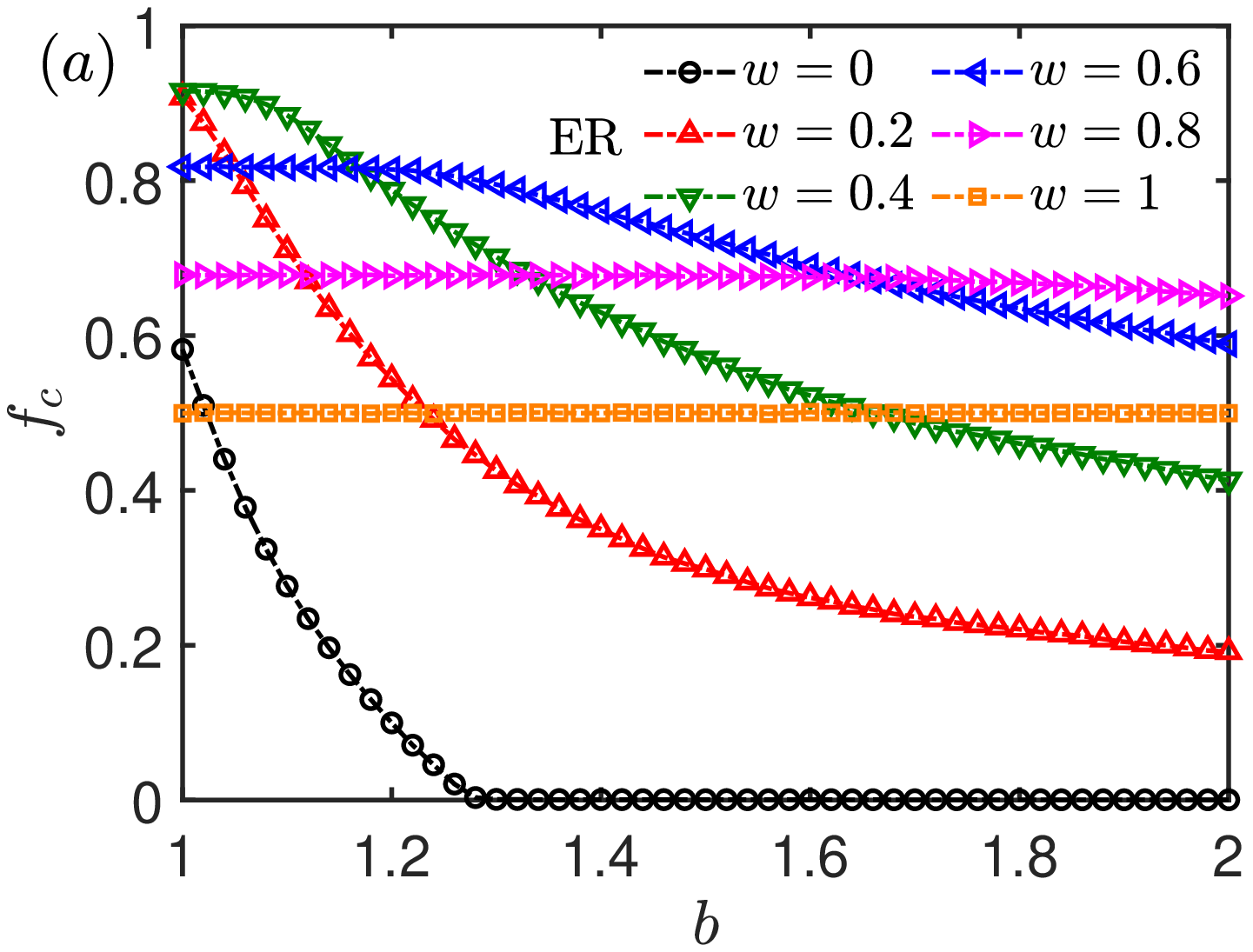}
	\includegraphics[width=0.45\linewidth]{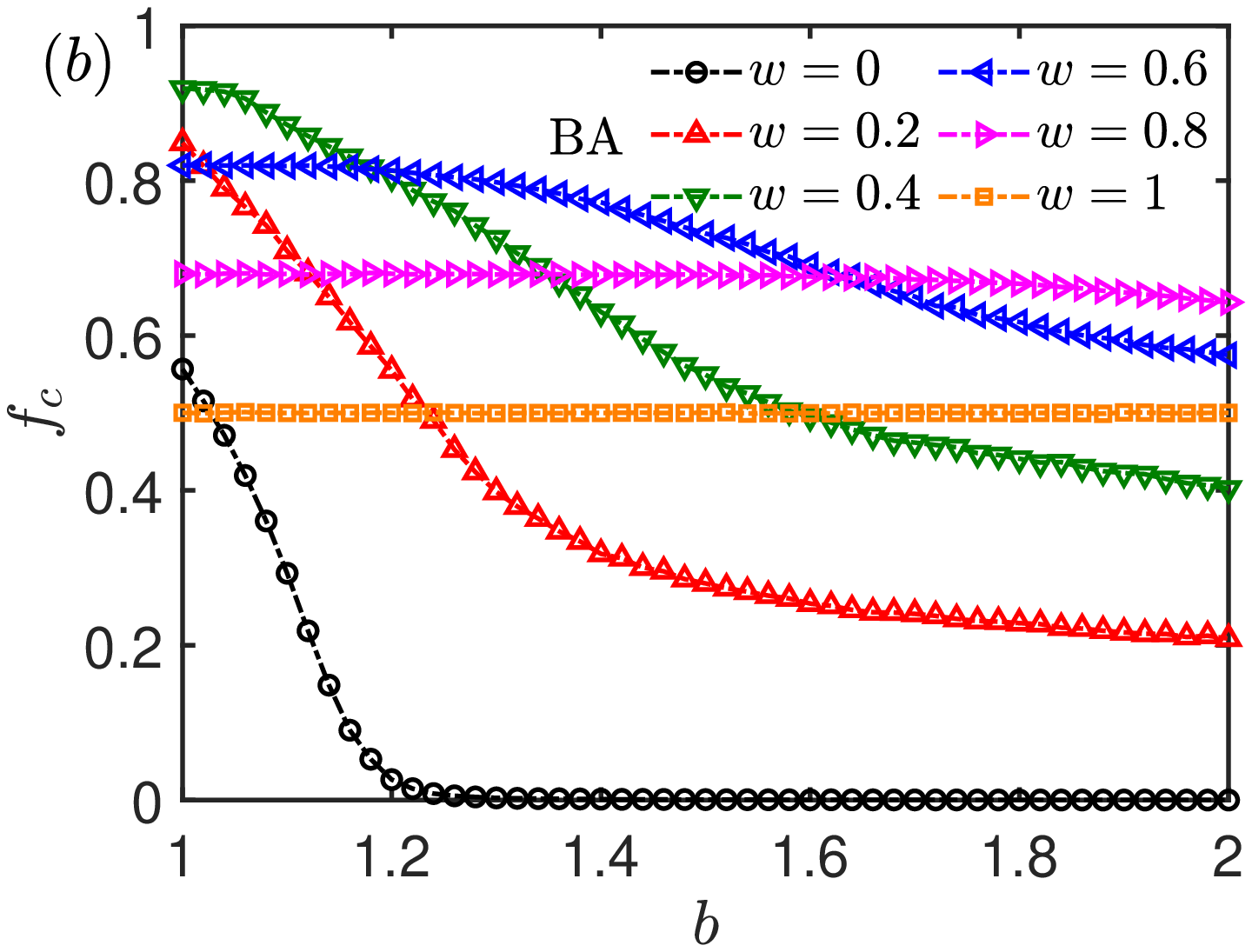}\\
	\includegraphics[width=0.45\linewidth]{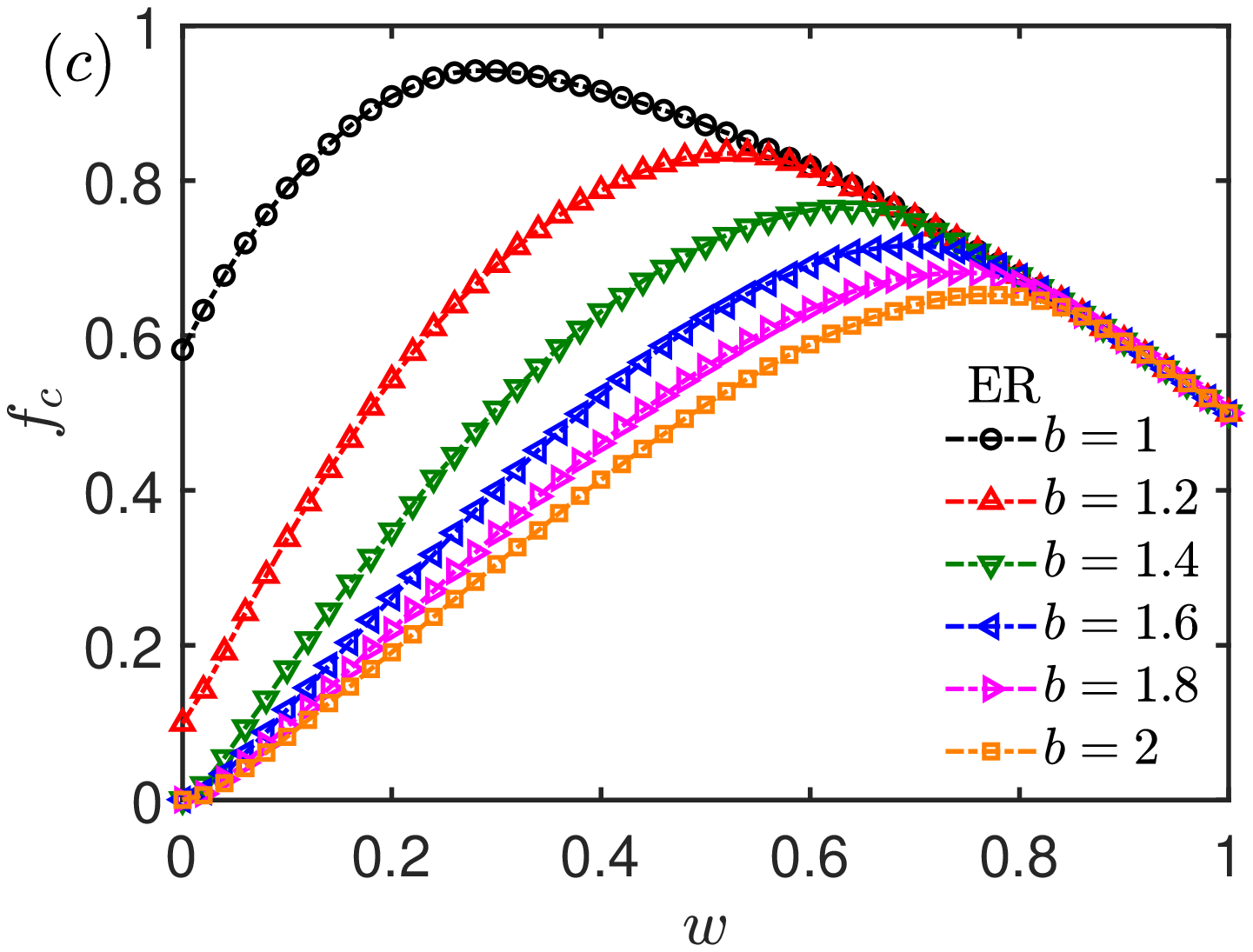}
	\includegraphics[width=0.45\linewidth]{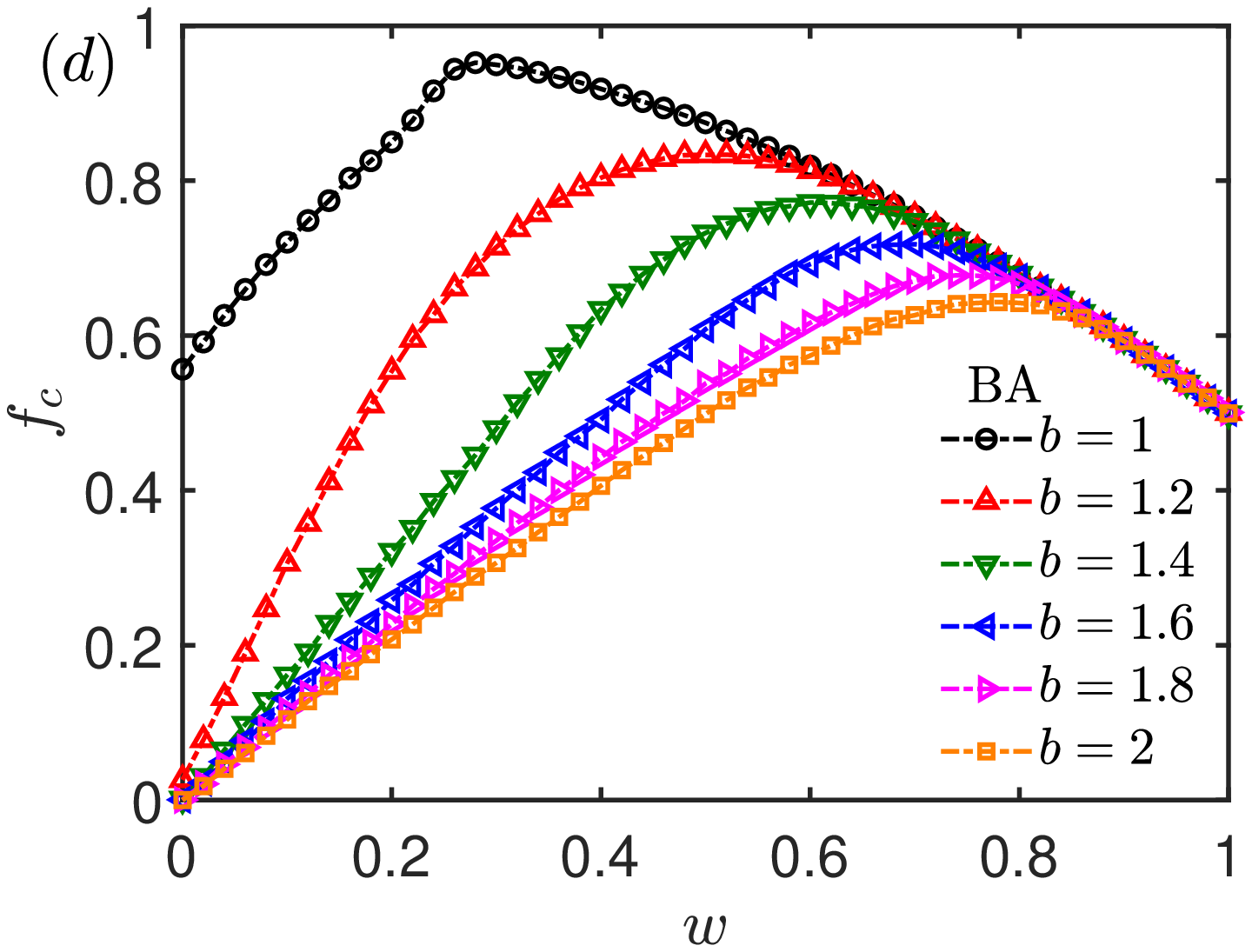}
	\caption{The impact of underlying structures of the population.
		The dependence of cooperation prevalence $f_c$ on two parameters with SM on ER networks (a,c) and BA networks (b,d).
		Parameters: $N=2^{20}$, the average degree $\langle k \rangle=4$.
	}
	\label{Fig:ERBAJGpf}
\end{figure*}

The results of structural mixing are shown in Fig.~\ref{Fig:ERBAJGpf}, where the TFT-players are randomly chosen. Fig.~\ref{Fig:ERBAJGpf}(a,b) show the cooperation prevalence $f_c$ as a function of temptation $b$ for different $\omega$ in ER and BA networks, respectively. When $\omega=0$, the results are consistent with the previous study. Note that, because the evolution is based upon the average payoffs rather than the total payoffs, the value of $f_c$ for BA networks is not higher than that of ER case as might be expected ~\cite{santos2005scale,wu2007evolutionary}. In both cases, the inclusion of TFT-players promotes cooperation, and increasing the temptation $b$ generally decreases $f_c$ except for the case of $\omega\rightarrow 1$.
Fig~\ref{Fig:ERBAJGpf}(c,d) show the dependence on the probability $\omega$, and the existence of an optimal $\omega_o$ is clearly seen that yields the best cooperation. Compare to the lattice case (Fig.~\ref{Fig:JGpfp}(a,b)), complex network topologies do not alter the dependence on the mode mixing qualitatively.

To further investigate the impact of the network heterogeneity, the following three ways are adopted to select TFT-players:

i) \emph{Neutral correlation} --- $\omega N$ players are randomly chosen irrespective of their degrees; this is the way we used in Fig.~\ref{Fig:ERBAJGpf}. 

ii) \emph{Positive correlation} ---  nodes with larger degrees are chosen to use the TFT rule; 

iii) \emph{Negative correlation} --- nodes with smaller degrees are selected. 

Specifically, in the latter two ways, we rank all nodes based on their degrees in a descending/ascending order, and pick the first $\omega N$ nodes. In such a way, the selection of TFT-players has a positive/negative correlation with their degrees. 

\begin{figure}[!h]
	\centering
	\includegraphics[width=0.45\linewidth]{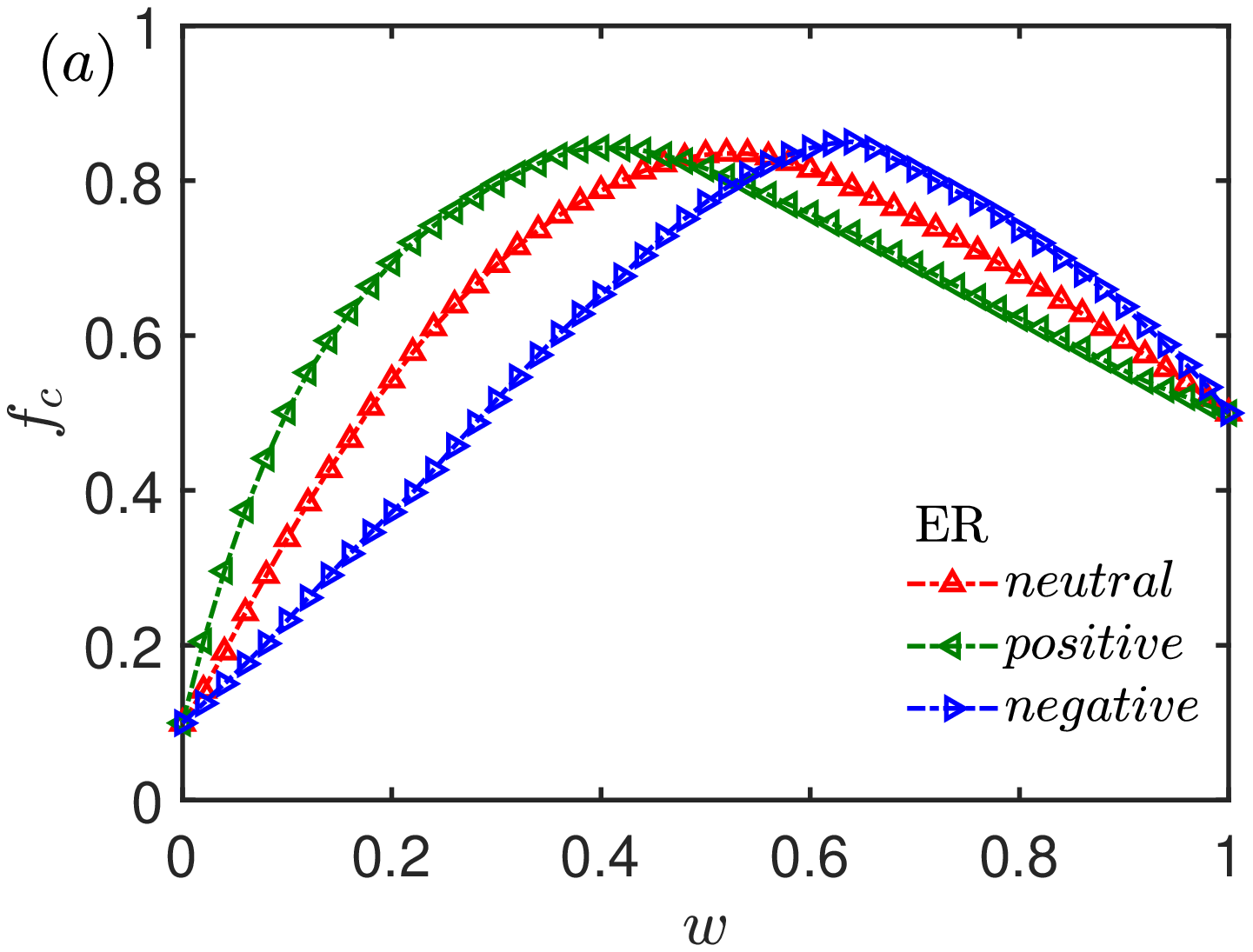}
	\includegraphics[width=0.45\linewidth]{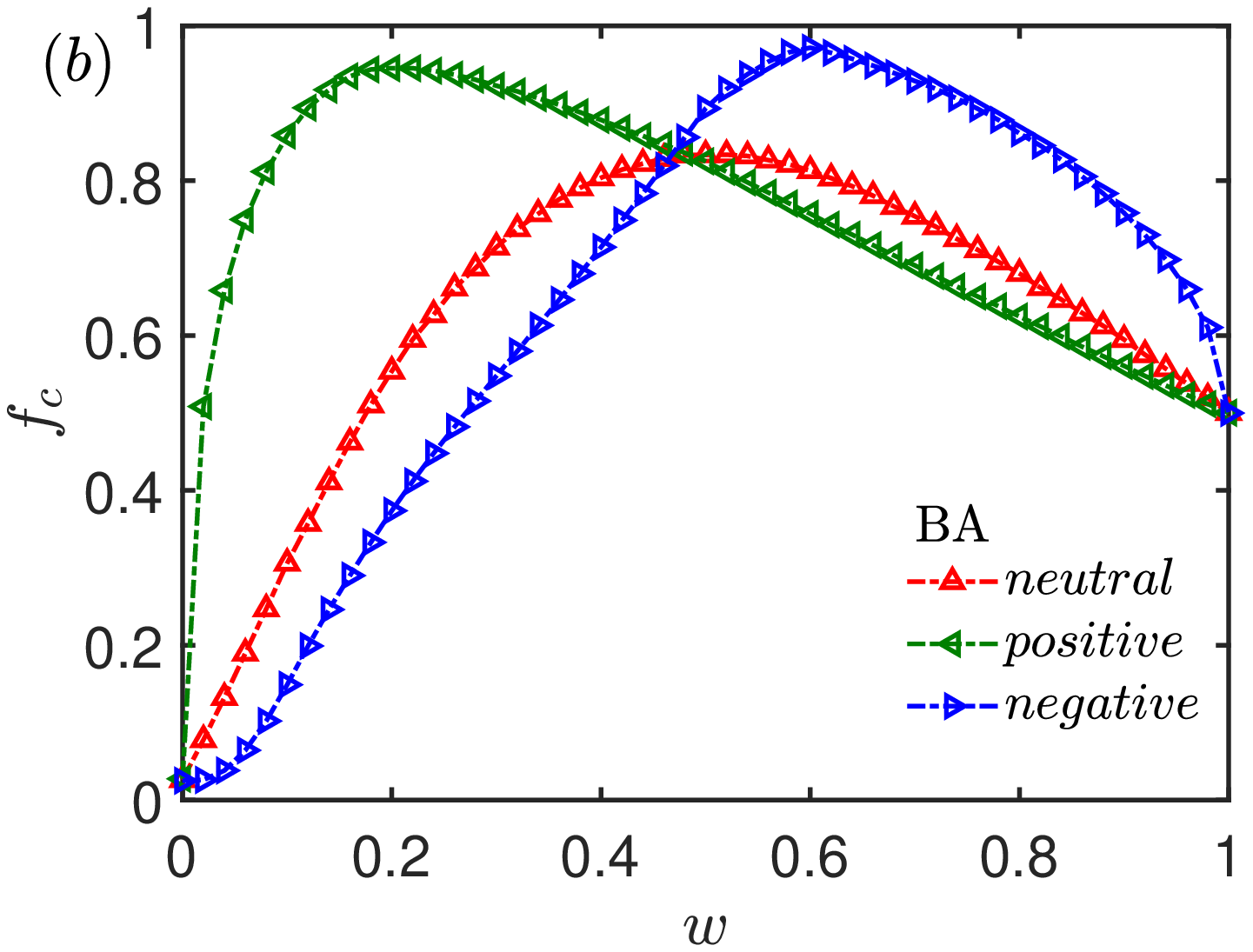}
	\caption{The impact of the degree correlation.
		The evolution of cooperation with SM for three different ways of selecting TFT-players on (a) ER and (b) BA networks.
		Parameters: $N=2^{20}$, the average degree $\langle k \rangle=4$, $b=1.2$.
	}
	\label{Fig:ERBAJGfcw}
\end{figure}

The impact of different correlations on both networks is shown in Fig.~\ref{Fig:ERBAJGfcw}. We see that in both cases the positive correlation shifts the optimal ratio $\omega_o$ to be smaller, and in the opposite direction for the negative correlations. Besides, this shift is more pronounced in BA networks. 
The reason lies in the approximate correspondence between the cooperation prevalence and the number of interactions of Fermi-TFT type. 
When hub nodes are occupied by TFT-players, a smaller fraction is just able to have the most Fermi-TFT interactions. By contrast, when the TFT-players are periphery nodes, a more fraction is needed.   

The probabilistic mixing is also implemented on both networks (see Sec. II in  Supplemental Material), the results are quite similar to the lattice case (Fig.~\ref{Fig:GLpfp}(a,b)). 

\subsection{Theoretical analysis} 
\begin{figure}[!h]
	\centering 
	\includegraphics[width=0.6\linewidth]{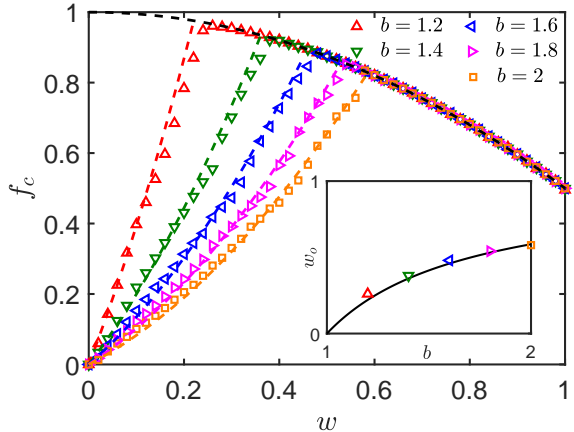}
	\caption{Results from the mean-field theory. The comparison between the theoretical analysis results (dashed lines) and the numerical simulation results (scatters) on well-mixed network for SM, where $\omega$ is the mixing ratio.
		The inset shows the dependence of optimal mixing ratio $\omega_o$ on $b$, the solid line is given by Eq.~(\ref{eq:wo}).
		Other parameter: $N=2^{9}$. 
	}
	\label{Fig:Theoretical}
\end{figure}

To understand why the bimodality is able to promote the cooperation prevalence, we provide a mean-field analysis for the structural mixing implementation. For simplicity, let's consider a well-mixed population, numerical results are shown in Fig.~\ref{Fig:Theoretical}, where qualitatively the same observations are seen compared to the above findings (e.g. Fig.~\ref{Fig:JGpfp}(b) and Fig.~\ref{Fig:ERBAJGpf}(c,d)). 

The four types of edge strategies are denoted as $C^T$, $D^T$, $C^F$, $D^F$,  where $T$ (TFT) and $F$ (Fermi) represent the type of player holding that strategy. All interaction pairs are summarized in the following fraction matrix: 
\begin{equation} \label{eq:matrix1}
	\bordermatrix{
		&C^T&D^T&C^F&D^F\cr
		C^T&f_{11}&f_{12}&f_{13}&f_{14}\cr
		D^T&f_{21}&f_{22}&f_{23}&f_{24}\cr
		C^F&f_{31}&f_{32}&f_{33}&f_{34}\cr
		D^F&f_{41}&f_{42}&f_{43}&f_{44}
	}.
\end{equation}
Each item represents the fraction of the associated two edge strategies in the population, which can also be interpreted as the probability of finding that edge. For example, $f_{14}$ is the probability of finding the links that connect a TFT-player with C and a Fermi-player holding the strategy D. Since these probabilities are irrespective of the pair order, therefore 
\begin{equation} \label{eq:fij}
	f_{ij}=f_{ji}, \hspace{1cm}  (i,j=1,2,3,4).
\end{equation}

The sum of each row is the overall strategy density respectively for TFT and Fermi rule, defined as below, 
\begin{equation} \label{eq:y11y12y13y14}
	f_{11}+f_{12}+f_{13}+f_{14}=f_{C^T}, 
\end{equation}
\begin{equation} \label{eq:y21y22y23y24} 
	f_{21}+f_{22}+f_{23}+f_{24}=f_{D^T}, 
\end{equation}
\begin{equation} \label{eq:y31y32y33y34}
	f_{31}+f_{32}+f_{33}+f_{34}=f_{C^F},
\end{equation}
\begin{equation} \label{eq:y41y42y43y44}
	f_{41}+f_{42}+f_{43}+f_{44}=f_{D^F}.
\end{equation}
The four fractions satisfy the following relations in the structural mixing, 
\begin{equation} \label{eq:fT}
	f_{C^T}+f_{D^T}=\omega,
\end{equation} 
\begin{equation} \label{eq:fF}
	f_{C^F}+f_{D^F}=1-\omega.
\end{equation} 

Now let's consider all three type of interactions. 
For type i) interactions, which correspond to the Fermi-Fermi pairs, i.e., the lower right corner of the matrix, the four items can be expressed in the mean-field sense as 
\begin{equation} \label{eq:type1a}
	f_{33}=f^2_{C^F}, f_{44}=f^2_{D^F}, 
\end{equation}
\begin{equation} \label{eq:type1b}
	f_{34}=f_{43}=f_{C^F}f_{D^F}.
\end{equation}

For type ii) interactions, where TFT-player encounters TFT-player, these edge strategies will finally evolve into C-C and D-D pairs with an equal chance, meaning
\begin{equation} \label{eq:TFT_diagonal}
	f_{11}=f_{22}=\omega ^2/2, 
\end{equation}
\begin{equation} \label{eq:TFT_nondiagonal}
	f_{12}=f_{21}=0. 
\end{equation}

The cooperation of TFT-players $f_{C^T}$ comes from two contributions, one is from the TFT-TFT interactions $f_{11}$; The other is from type iii) interactions, where TFT-players copy exactly what Fermi-players did to them. This then leads to 
\begin{equation} \label{eq:CTandCF}
	f_{C^T}=\omega f_{C^F} + \omega ^ 2/2.
\end{equation}
Inserting Eq.~(\ref{eq:type1a}-\ref{eq:CTandCF}) into Eq.~(\ref{eq:y11y12y13y14},~\ref{eq:y31y32y33y34}), we have 
\begin{equation} \label{eq:y13y14}
	f_{13}+f_{14}=\omega f_{C^F},
\end{equation}
\begin{equation} \label{eq:y31y32}
	f_{31}+f_{32}=\omega f_{C^F}.
\end{equation}
With Eq.~(\ref{eq:fij}),	 we find 
\begin{equation}\label{eq:y14y32y41y23}
	f_{14}=f_{32}=f_{41}=f_{23}=\omega f_{C^F}-f_{31}.
\end{equation}

With these relationship, the matrix (\ref{eq:matrix1}) can be rewritten as follows
\begin{equation} \label{eq:matrix2}
	\begin{pmatrix}
		\omega^{2}/2&0&f_{31}&\omega f_{C^F}-f_{31}\\
		0&\omega^{2}/2&\omega f_{C^F}-f_{31}&\omega f_{D^F}-\omega f_{C^F}+f_{31}\\
		f_{31}&\omega f_{C^F}-f_{31}&f_{C^F} f_{C^F}&f_{C^F} f_{D^F}\\
		\omega f_{C^F}-f_{31}&\omega f_{D^F}-\omega f_{C^F}+f_{31}& f_{D^F} f_{C^F}&f_{D^F} f_{D^F}
	\end{pmatrix}
\end{equation}
where $f_{D^F}=1-\omega-f_{C^F}$ from Eq.~(\ref{eq:fF}).

Through numerical simulations, we identify the following relationships when $\omega\le\omega_o$ (before the absorbing state is reached $s_i<1$ for the Fermi-players)
\begin{equation} \label{eq:payoff1}
	\Pi_{C^F}=\Pi_{D^F}=\Pi_{F}
\end{equation}
\begin{equation} \label{eq:payoff2}
	\Pi_{F}=\Pi'_{T},
\end{equation}
where the payoff of $D^T-D^T$ is excluded in $\Pi'_{T}$. Specifically,
\begin{equation} \label{eq:payoffmatrix1}
	\small
	\frac{f_{31}R+f_{32}S+f_{33}R+f_{34}S}{f_{C^F}}=\frac{f_{41}T+f_{42}P+f_{43}T+f_{44}P}{f_{D^F}},
	\normalsize
\end{equation}
\begin{equation}
	\begin{split} \label{eq:payoffmatrix2} 
		\small
		&\frac{f_{31}R+f_{32}S+f_{33}R+f_{34}S+f_{41}T+f_{42}P+f_{43}T+f_{44}P}{1-\omega}\\
		&=\frac{f_{11}R+f_{12}S+f_{13}R+f_{14}S+f_{21}T+f_{23}T+f_{24}P}{\omega-f_{22}}.
	\end{split}
\end{equation}
\normalsize
By combining the elements in matrix (\ref{eq:matrix2}) and we solve Eq.~(\ref{eq:payoffmatrix1}), Eq.~(\ref{eq:payoffmatrix2}), we get
\begin{equation} \label{eq:fCF}
	f_{C^F}=\frac{\omega}{\sqrt{2}(b-1)}.
\end{equation}
Insert the above expression into Eq.~(\ref{eq:CTandCF}), we obtain $f_{C^T}$. Finally, we add them up $f_{C}=f_{C^T}+f_{C^F}$, the expression of overall cooperation prevalence is
\begin{equation} \label{eq:fC}
	f_{C}=\frac{(1+\omega) \omega}{\sqrt{2}(b-1)}+\frac{\omega^2}{2}.
\end{equation} 
Together with Eq.~(\ref{eq:fCF}), our analysis show that the fraction of cooperation for Fermi-players $f_{C^F}$ increases linearly with the fraction of TFT-players. Without TFT-players ($\omega=0$), the cooperation disappears, as is well-known in previous studies~\cite{taylor1978evolutionary,Maynard1982Evolution}. The overall cooperation prevalence $f_C$ is quadratic function of $\omega$.

Since $f_{C^F}\le1-\omega$ given by the Eq.~(\ref{eq:fF}), which leads to 
\begin{equation} \label{eq:wo}
	\omega\le\frac{\sqrt{2}(b-1)}{\sqrt{2}(b-1)+1}\equiv\omega_o.
\end{equation} 
When $\omega=\omega_o$, full cooperation is reached for Fermi-players, also with the least D-D pairs for type ii) interactions. When $\omega>\omega_o$, the overall cooperation prevalence $f_C$ satisfies Eq.~(\ref{eq:fc_approx_right}).   

The comparison between theoretical analysis results (Eq.~(\ref{eq:fC}) for $\omega\le\omega_o$, and Eq.~(\ref{eq:fc_approx_right}) for $\omega>\omega_o$) and numerical results is shown in Fig.~\ref{Fig:Theoretical}, confirming the correctness of the above derivation. The inset shows the dependence of $\omega_o$ on the parameter $b$, the value shifts to be larger with increasing $b$, in line with the numerical observations.

\section{Conclusion \& discussion}
In summary, motivated by the diversity of human behavioural modes in the realistic evolution of human cooperation, we build a model with a mixture of Fermi and TFT rules, two commonly seen modes in previous studies. In the first implementation where the individuals use one fixed rule, we find that the mode mixing can promote cooperation, and there exists an optimal amount of TFT individuals that bring the highest level of cooperation. In the second implementation, individuals probabilistically behave in the either mode, the full cooperation is always achieved if the probability of using the TFT rule is beyond a critical value in the mixing population. These findings are verified on two complex networks, where the degree heterogeneity only changes the results quantitatively. Finally, we derive a semi-analytic mean-field treatment for the first implementation, give the dependence of cooperation prevalence on the mixing ratio and the game parameters, explicitly revealing how the mixture of Fermi and TFT rules promotes cooperation.   

Note that, even though the extension of the Fermi updating rule~\cite{szabo1998evolutionary,szabo2005phase} from the node-based strategy to the edge-based version does not alter the cooperation prevalence at all, this extension is necessary in our study for the need of mixture since the TFT is also via an edge-based updating scheme. Furthermore, the edge-based strategy scheme seems more reasonable in most realistic scenarios, since individuals treat their different neighbors potentially in different strategies, not a uniform strategy against all their neighbors as most current game-theoretical models assume. 
Besides, when the Fermi rule is replaced with the deterministic follow-the-best rule~\cite{nowak1992evolutionary}, the above findings remain unchanged qualitatively. Other model variants like the replacement of asynchronous updating with the synchronous scheme also show similar observations.

While the Fermi-rule is regarded as the typical imitation rule in the strategy updating, it's not easy to interpret TFT rule as any single attribute or label. As the winning strategy in Robert Axelrod' two tournaments, TFT is generally considered as being strategic for its clear, nice, provocable, and forgiving properties~\cite{Axelrod1981The} and is suggested to be the cooperation mechanism in some animal communities. Therefore, we would rather not to make specific interpretation of the two mode mixture. In brief, our study shows that the incorporation of behavioral bimodality help explain the emergence of cooperation. Together with the related works~\cite{Szolnoki2015Conformity,Szolnoki2018Competition,Danku2018Imitate,Amaral2018Heterogeneous,Masuda2012evolution,Zheng2022probabilistic,Zheng2022pinning}, the research avenue of considering different behavioral modes might be indispensable for modeling human behaviors in many realistic scenarios.

\section*{Acknowledgments}
We are supported by the Natural Science Foundation of China under Grants No. 12075144 and 12165014. We thank Prof. Weiran Cai (Soochow University) for helpful comments.

\bibliographystyle{elsarticle-num-names} 
\bibliography{fmtftenglish}









\end{document}